\documentclass[twocolumn,final,number,sort&compress]{elsarticle}
\usepackage[displaymath]{}
\usepackage{multirow}
\usepackage{float}
\usepackage{amsmath}
\usepackage{amsfonts}
\usepackage{amssymb}
\usepackage{graphicx}
\usepackage[subrefformat=parens,labelformat=parens]{subfig}
\usepackage{natbib}
\usepackage{lipsum}
\usepackage{comment}
\usepackage[usenames]{color}
\usepackage{soul}
\usepackage{xurl}
\begin{document}

\definecolor{electricpurple}{rgb}{0.75, 0.0, 1.0}
\definecolor{electricgreen}{rgb}{0.1, 0.588, 0.378}
\newcommand{\blue}[1]{\textcolor{blue}{#1}} 
\newcommand{\green}[1]{\textcolor{electricgreen}{#1}} 
\newcommand{\purple}[1]{\textcolor{electricpurple}{#1}} 

\begin{frontmatter}

\title{Implementation and validation of realistic (n,x) reaction yields in GEANT4 utilizing a detailed evaluated nuclear reaction library below 20~MeV}

\author[CMU,JINA]{P. Tsintari\corref{cor1}} 
\ead{tsint1p@cmich.edu}
\author[CMU,JINA]{G. Perdikakis\corref{cor1}}
\ead{perdi1g@cmich.edu}
\author[LANL]{H.Y. Lee\corref{cor1}} 
\ead{hylee@lanl.gov}
\author[LANL]{S.A. Kuvin}
\author[UBerkeley,LBNL]{A. Georgiadou}
\author[KAERI]{H.I. Kim}
\author[LANL]{D. Votaw}
\author[ONL]{L. Zavorka}

\address[CMU]{Department of Physics, Central Michigan University, Mt. Pleasant, MI 48859, USA}
\address[LANL]{Los Alamos National Laboratory, Los Alamos, NM 87545, USA}
\address[JINA]{Joint Institute of Nuclear Astrophysics - Center for the Evolution of the Elements, Michigan State University, East Lansing, MI 48824, USA}
\address[UBerkeley]{Department of Nuclear Engineering, University of California, Berkeley, CA 94720, USA }
\address[LBNL]{Nuclear Science Division, Lawrence Berkeley National Laboratory, Berkeley, California 94720, USA
}
\address[KAERI]{Nuclear Physics Application Research Division, Korea Atomic Energy Research Institute, Yuseong-gu, Daejeon, Korea}
\address[ONL]{Oak Ridge National Laboratory, Oak Ridge, TN 37830, USA}

\cortext[cor1]{Corresponding authors}

\begin{abstract}

Neutron-induced reactions with charged particle emission play an important role in a variety of research fields ranging from fundamental nuclear physics and nuclear astrophysics to applications of nuclear technologies to energy production and material science. Recently, the capability to study reactions with radioactive targets has become important to significantly advance research in explosive nucleosynthesis and nuclear applications. To achieve the relevant research goals and study (n,x) reactions over a broad neutron beam energy range,  the Low Energy Neutron-induced charged-particle (Z) chamber (LENZ) at Los Alamos Neutron Science Center (LANSCE) was developed along with varied ancillary instrumentation to enable the aforementioned research program. For the (n,x) reactions of interest at low energies, a precise simulation of the discrete spectrum of emitted charged particles is essential. In addition, since LANSCE is a user facility, a simulation application that can be easily accessible by users has high value. With these goals in mind, we have developed a detailed simulation using the GEANT4 toolkit. In this work, we present the implementation and the validation of the simulation using experimental data from recent campaigns with the LENZ instrument. Specifically, we benchmark the simulation against a similar MCNP-based tool and determine the realistic range of applicability for the probability biasing technique used. We describe our implementation of an evaluated library with angular distribution and partial cross-section data, 
and we perform a validation of the application based on comparisons of simulated spectra with the experimental ones, for a number of targets used in previous experimental campaigns. Last, we discuss the limitations, caveats, and assets of the simulation code and techniques used.


\end{abstract}
\end{frontmatter}

\section{Introduction}
\label{sec:Introduction}

Neutron-induced reactions are of great importance in nuclear astrophysics since the bulk of elements heavier than iron are synthesized in the cosmos by either of two nucleosynthesis processes involving neutron capture, namely the slow neutron capture (s-process) or the rapid neutron capture (r-process) process \cite{b2fh, Cam57}. However, these two processes are not adequate to explain all the observed abundances. It is known that to explain the production of the few stable neutron-deficient isotopes \cite{Wal97a}, shielded from the r- and s- processes (p-nuclei), new processes have to be invoked (i-process, p-process or $\gamma$-process). Advances in abundance observations in metal-poor stars and in particular of the elements Sr, Y, and Zr suggest that a previously unexplored nucleosynthesis process should exist (in addition to the s-, r-, and $\gamma$- process) \cite{Fre05a}. This process has been called the LEPP (Lighter Element Primary Process) \cite{Qia07a, Mon07a} but the precise mechanism by which it synthesizes heavy elements still eludes us. 

A strong candidate for the LEPP that involves neutron-induced reactions motivates this work. The neutrino-p ($\nu$p)~process \cite{Fro06a} could play the role of LEPP at the slightly proton-rich regions in the neutrino driven wind of core-collapse supernovae \cite{Arc11a}, as predicted by long-term hydrodynamics simulations with sophisticated neutrino transport. 

We provide here a brief description of the relevant nucleosynthesis scenario. Hydrodynamic simulations indicate that the neutrino-driven wind could be slightly proton-rich during the early phases of the wind evolution after a type II supernova explosion \cite{liebend03,buras06b,cf06a,fischer10,huedepohl10}. In such conditions, a nucleosynthesis path along the proton-rich side of the chart is favored.  The strong neutrino and anti-neutrino fluxes from the proto-neutron star, however, allow for electron anti-neutrino absorption on protons ($\overline{\nu}_\mathrm{e}+\mathrm{p}\rightarrow \mathrm{n}+\mathrm{e}^+$) to produce a small residual neutron density.

This small density of neutrons prevents the typical reaction flow slow down that would happen as soon as the nucleosynthesis path crosses nuclei with relatively long $\beta$-decay half-lives such as $^{56}$Ni and $^{64}$Ge and marks the beginning of the $\nu$p process. 

The availability of neutrons allows for (n,p) reactions to happen as temperatures drop between 3~GK and 1.5~GK. The (n,p) reactions are faster than the $\beta$-decays, and hence bypass them. This bypass of the flow bottleneck allows the synthesis of heavy elements beyond iron \cite{Fro06a, pruetII, wanajo06,  Wan11a, aacf}. 

Attempts to quantify the importance of the $\nu$p process to nucleosynthesis are inhibited, among other factors, by the lack of experimental data on the bottleneck nuclei ($^{56}$Ni, and others such as $^{64}$Ge)and their (n,p) reaction rates. Currently, theoretical predictions of these rates come with significant uncertainties related to the details of the optical potential away from stability. 

Recently, the importance of certain (n,p) and (n,a) reactions to questions related to the radiogenic heating of terrestrial planets with implications for the habitability of exoplanets was explored by the group at Central Michigan University \cite{Gastis2020a}.

Beyond the nuclear astrophysics points discussed above, neutrons are an essential probe in nuclear physics studies aiming to investigate nuclear reaction mechanisms and extract nuclear structure information \cite{MAZZONE201833}. Neutron-induced reactions with charged particle emission are significant for a variety of research fields ranging from fundamental nuclear physics and astrophysics to applications of nuclear technologies for energy production, nuclear medicine, and material analysis. 

An example of an issue, typically encountered in nuclear engineering, is related to the development of fusion and fission reactors. The generation of charged particles through fast-neutron induced reactions such as (n,p), (n,$\alpha$), (n,n'p), and (n,n'$\alpha$) leads to the formation of hydrogen and helium gases during reactor operations. These reactions are essential for estimating residual radioactivity and resolving licensing and radioactive waste generation issues. These reactions are also related to the radiation damage inflicted in reactors through the introduction of gas-filled pores into the reactor's structural materials \cite{PhysRevC.85.024624}.

A review of existing literature on neutron-induced reaction cross-sections reveals a lack of experimental data sets on (n,p), (n,$\alpha$) reactions for medium and high Z isotopes over a broad neutron energy range. Data sets are often discrepant by factors of a few at 14~MeV, and the majority of experimental data sets are not covering a broad enough neutron-energy range to meet the requirements for nuclear astrophysics or the other nuclear applications mentioned above.

For astrophysics applications, evaluated libraries do not cover energies below 1~MeV, which correspond to the stellar temperatures of interest for explosive nucleosynthesis. For applications in nuclear engineering, data on product angular distributions are missing and the available experimental cross-section data with broad energy coverage are too sparse to support reliable nuclear data evaluations. 

Research efforts to address the lack of experimentally constrained (n,p) and (p,n) reaction rates for short-lived nuclei by developing an appropriate technique for (p,n) reactions using FRIB radioactive beams are under way since 2017 and have culminated in the demonstration of the technique using the $^{40}$Ar(p,n)$^{40}$K reaction in inverse kinematics \cite{Gastis2021}. The technique is currently undergoing commissioning at the SECAR recoil separator. The first radioactive beam experiment with the new technique will be the measurement of the $^{56}$Co(p,n)$^{56}$Ni reaction, opening the road for experiments with more exotic species. The measurement at LANL of the $^{56}$Ni(n,p)$^{56}$Co reaction, besides the aforementioned scientific importance  will also provide an important benchmark for this effort as well.

Historically, the majority of neutron cross-section measurements have been performed using neutron generators at neutron energies around 14~MeV, or with small accelerator-based sources like the $^2$H(d,n)$^3$He reaction for neutron energies between 3-10~MeV, and $^7$Li(p,n)$^7$Be for energies below 1~MeV. Many of these measurements have traditionally used the neutron activation technique that (while providing high statistics) does not typically retain information about short-lived excited state production or light particle angular distributions.     

Dataset shortcomings such as these, especially in the case of medium-Z nuclei, hinder subsequent theoretical analyses and determinations of the appropriate statistical model parameters used in the Hauser Feshbach theory. Therefore, the systematic uncertainties of data evaluations and nucleosynthesis calculations increases.

The Low Energy Neutron-induced charged-particle (Z) chamber (LENZ)~\cite{LeePSD2019} was developed at Los Alamos Neutron Science Center (LANSCE) to provide the high-quality nuclear data needed for modern applications. It is designed for the study of neutron-induced charged particle reactions with neutrons in the energy range from thermal to several tens of MeV, and is optimized for measurements of differential cross-sections and angular distributions with a small energy step over the broad-energy neutron beam at LANSCE, and with a total geometrical efficiency surpassing previous systems in operation \cite{NZhaight, Devlin2009, Kunieda2012, osti_10165734}.

For the nuclear astrophysics applications that motivate this work, the LENZ instrument has to fulfill additional requirements. The instrument has to operate successfully in a high-radiation field generated by the radioactive reaction target of interest. For reactions with medium-Z nuclei, like $^{56}$Ni, the amount of radioactive emission is such that dictates a thorough understanding of all possible beam-induced background sources in the experiment. The primary goal of this work is to develop a Monte Carlo simulation that can be utilized by users of the LANSCE facility in order to understand sources of background, estimate experimental yields and interpret measurement data. The simulation uses GEANT4 \cite{AGOSTINELLI2003250}, an open source Monte Carlo simulation toolkit, made available by CERN, designed for modeling interactions of particles with matter. It has proven to be very versatile for the description of complex geometries with multiple materials and is shown to be reliably accurate in the description of neutron scattering processes \cite{ZUGEC201457} and neutron detector response simulations \cite{GARCIA201773}. To ensure the realistic reproduction of (n,x) reaction yields and particularly the low excitation energy spectra of emitted protons and alphas we have implemented a state of the art evaluated library \cite{KIM2020163699} that includes detailed partial cross sections and angular distributions for up to the first forty levels of excitation of the residual nuclei studied. To assess the validity of the simulated spectra we compared our GEANT4 simulations with experimental data on (n,p) and (n,a) reactions acquired during a recent experimental campaign. 
The manuscript is organized as follows: In section \ref{sec:Experimental_setup} we describe the LENZ experimental configuration, including the characteristics of the LANSCE neutron beam as used in the simulation. In section \ref{sec:Geant}, we describe the procedure to model the LENZ setup using the GEANT4 toolkit, and discuss the implementation of the physics aspects of the simulation, including the description of particle emission to discrete levels. We discuss the results of the simulation and compare them with LENZ experimental data and an MCNP-based simulation in section \ref{sec:Results and Discussion}. The final conclusions from this work are presented in section \ref{sec:Conclusions}.

\section{Description of the experimental configuration}
\label{sec:Experimental_setup}

The simulations in this work revolved around the reproduction of the response of the LENZ instrument at Los Alamos National Laboratory (LANL). A dedicated instrument for the detection of charged particles, LENZ, is used for the study of neutron-induced reactions of the $(n,x)$ type. LENZ is making use of the white neutron beam generated from the spallation of high energy protons, at the Weapons Neutron Research (WNR) facility of LANSCE \cite{Lisowski1990, Lisowski2006}. The main aspects of the experimental setup used and simulated are described in the following sections.

\subsection{Fast neutron beam production at LANSCE}
\label{sec:Neutron_beam_LANSCE}

The WNR facility at LANSCE produces neutrons via the spallation process caused by 800~MeV protons impinging on an unmoderated tungsten target with an average proton current of approximately 4~$\mu$A. The proton beam is chopped and bunched before acceleration in order to give an adjustable pulse-to-pulse separation, typically of 1.8~$\mu$s. The pulse width approximately 125~ps (FWHM), allows for precise neutron Time Of Flight (TOF) measurements that are mainly limited by the timing resolution of the detector system. The ‘white’ neutron spectrum covers the energy range from about 100~keV to over 600~MeV and can be delivered simultaneously to six different flight paths \cite{LANSCE_fp}. The WNR broad-energy neutron spectrum and flux are ideal for studying double differential cross sections ($d^2\sigma / d\Omega dE_n$) for a variety of neutron-induced reactions. The flux and energy distribution of the neutron beam depend on each flight path's angle and distance from the tungsten spallation target. The neutron energy is typically deduced using the TOF method by measuring the time difference between the ``stop" signal produced by a neutron or other reaction particle at the detection location, and the time stamp of the ``start" signal produced by the 800 MeV proton beam via a pick-up circuit apparatus, right before hitting the tungsten spallation target. 

\begin{figure}[h]
\centering
\includegraphics[width=0.65\linewidth]{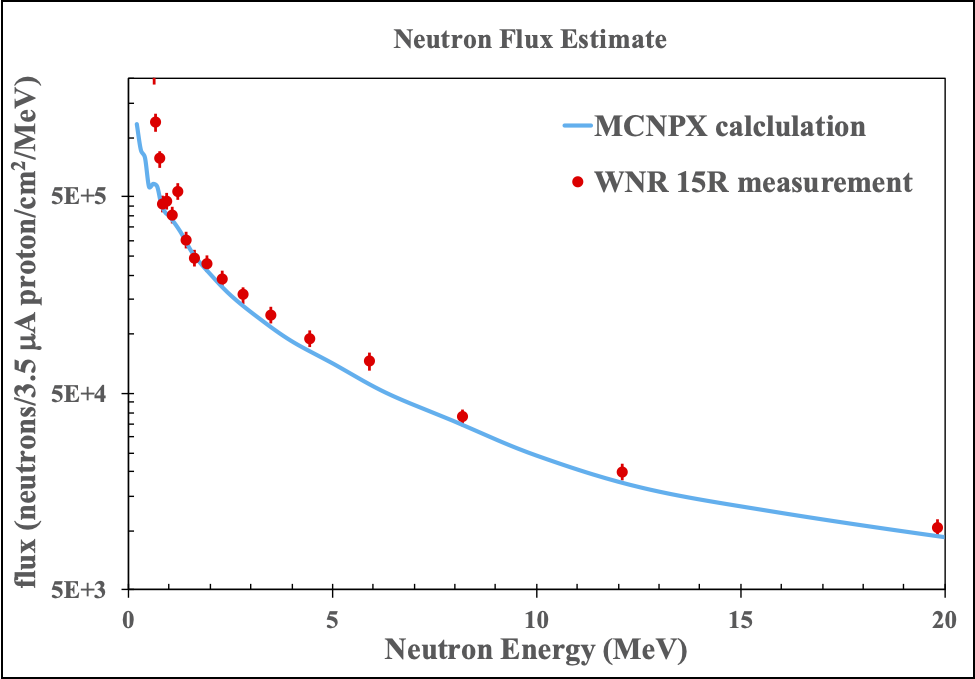}
\includegraphics[width=0.3\linewidth]{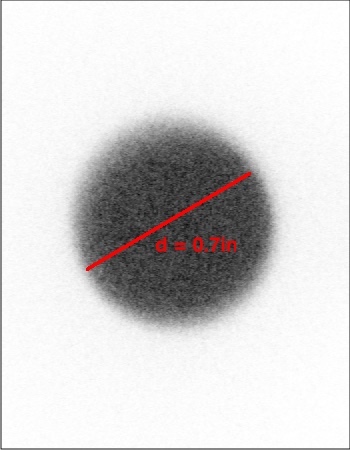}
\caption{(Left) The WNR neutron energy spectrum was simulated using MCNPX, which was benchmarked with the deduced neutron energy spectrum from the $^{238}$U(n,f) measurement~\cite{Lee_O16_2020, Kuvin_Cl35_2020}. (Right) The beam spot image at the WNR flight path 15R, where the LENZ experiments were performed. The diameter of the beam spot is 0.7 inches.}
\label{Beam}
\end{figure}

For the purposes of this work, the LENZ instrument was simulated placed at the 15$^{\circ}$R (Right) flight path at WNR. The detailed energy spectrum and flux of the neutron beam at 15R can be seen in Figure \ref{Beam}, where the points show the neutron flux measured using a $^{238}$U fission ionization chamber while an MCNP simulation currently in use by the facility \cite{Werner2017} is shown with a solid line. In order to provide the simulated neutron energy spectrum this full-scale MCNP calculation included the whole spallation neutron production process in the WNR target (T4). The MCNP result was benchmarked and found to be in good agreement with the $^{238}$U fission ionization flux monitor measurement, as shown in Figure \ref{Beam}(Left). Further details on the WNR neutron flux and spectrum measurements can be found in previous LENZ related publications \cite{Lee_O16_2020,Kuvin_Cl35_2020}.

\subsection{The Low Energy NZ (LENZ) setup}
\label{sec:LENZ_setup}

\begin{figure}[h!]
\centering
\includegraphics[width=0.9\linewidth]{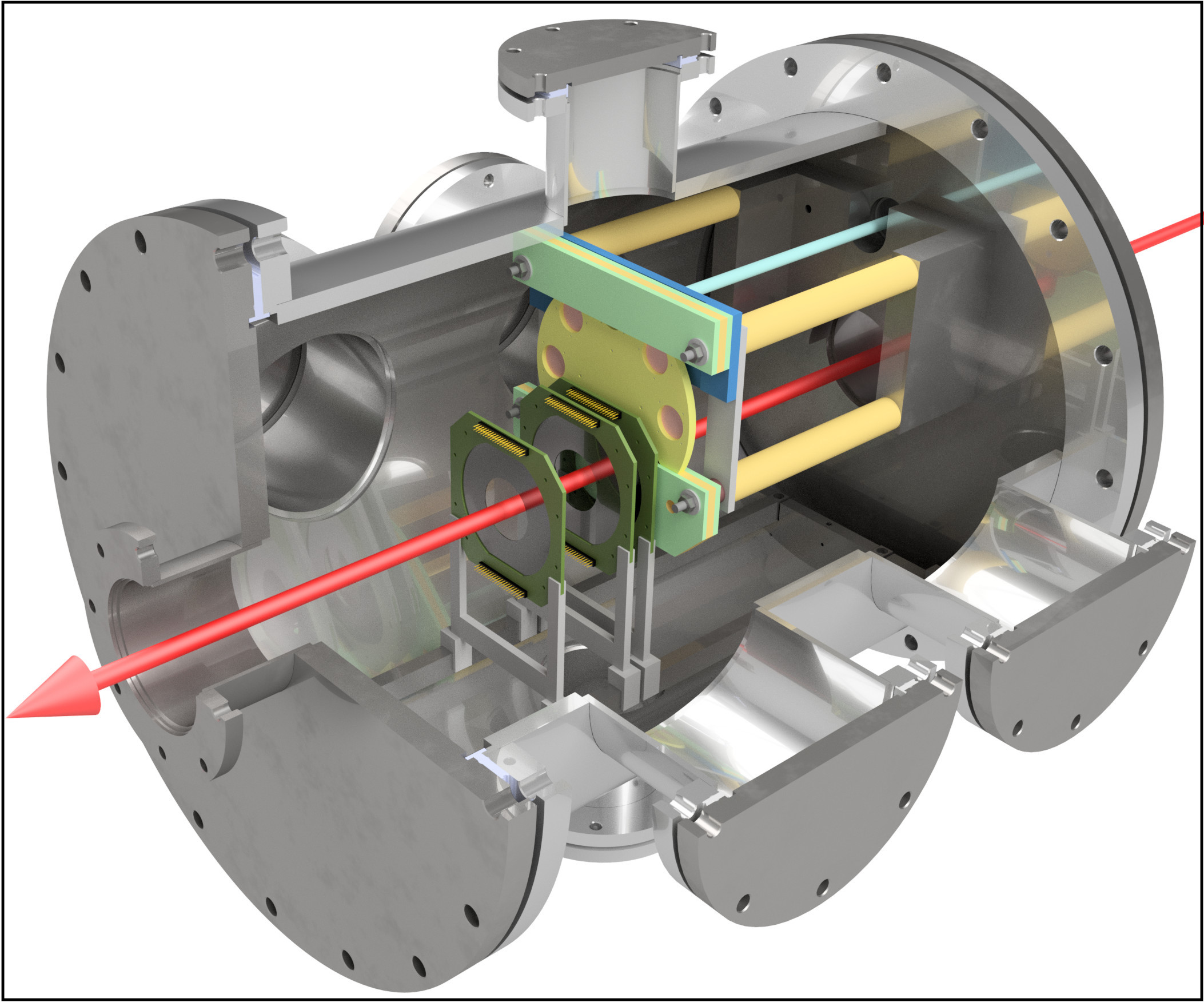}
\caption{Engineering design of the LENZ vacuum chamber; the cutout shows the detector setup with 3 position sensitive silicon detectors, and the target wheel. The red arrow represents the direction of neutron beams produced at LANSCE.}
\label{LENZ}
\includegraphics[width=0.5\linewidth]{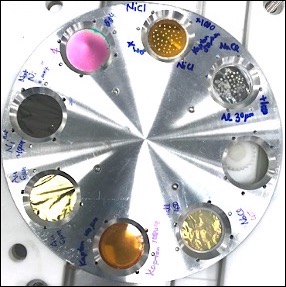}
\caption{Picture of the 8 position aluminum target wheel used in the middle of LENZ.}
\label{target-wheel}
\end{figure}

The LENZ instrument is a detector system for measuring neutron-induced reactions (n,x) that produce charged particles. It is designed to have a large solid angle coverage, a low detection threshold, and a timing resolution optimized for TOF measurements. LENZ was built to replace the preceding NZ chamber \cite{NZhaight, Devlin2009, Kunieda2012, osti_10165734} at WNR, which was used for the same purposes, but had a much more limited solid angle coverage. The experimental set-up consists of a cylindrical chamber, as shown in Figure \ref{LENZ}. The neutron beam enters and exits through two thin vacuum windows in opposite ends of the chamber, the material of which (typically Kapton and/or Aluminum) can be chosen according to the needs of each experiment. Having a minimum amount of material along the beam-path reduces the neutron-induced background in the detector system by minimizing neutron scattering, or neutron-proton recoil events in the case of hydrogen-rich window material such as Kapton.
 
The LENZ chamber is designed to be able to accommodate various different types of detectors, arranged along the beam axis. Examples of detectors that have been used in LENZ include annular double-sided silicon strip detectors (DSSD) such as S1 and S3 type ones \cite{Micron} of various thicknesses, a Frisch-gridded ionization chamber, and diamond detectors \cite{Cividec}. The positioning of the detectors can be configured according to detector type in any combination of detector thicknesses and in single or telescope $\Delta$E - E mode according to each experiment's needs. 

A reaction target wheel, shown in Figure \ref{target-wheel}, is placed at the center of the LENZ chamber and consists of an aluminum disk with 8 positions to hold different target materials. Each target position on the wheel has a diameter of 0.9~inches and can accommodate virtually any target of interest including backing materials, empty target frames, calibration sources, or reference targets for normalization purposes. To ensure the alignment of the targets to the neutron beam axis, the target wheel can rotate via a rotary feed-through mounted outside of the chamber, while maintaining vacuum in the chamber. The materials used and the geometry of the LENZ chamber are optimized to minimize background contributions from neutron scattering in the chamber, and to reduce any residual activation by favoring the production of short-lived isotopes in the prolonged runs that are typical of the facility. 

\subsection{Reaction targets studied in this work}
\label{sec:reactions_studied}

For the validation of the developed simulation and the physics processes involved we took advantage of the neutron induced reaction measurements performed with LENZ at LANSCE on selected target materials. In this work we included two standard backing materials for nuclear physics experiments, Mylar and Kapton, and a natural Ni target as an example of a multi-isotope target. The target materials are briefly described below and are listed along with the corresponding reactions studied in this work on Table~\ref{tab:list_reaction}.

The Mylar~(C$_{10}$H$_{8}$O$_{4}$) target foil had a thickness of 15~$\mu$m, while the Kapton (C$_{22}$H$_{10}$N$_2$O$_4$) target foil had a thickness of 100~$\mu$m. The Kapton foil of this commercially available thickness is not suitable for optimized charged particle detection but it was used to test the yield from a thick kapton backing. A $^{nat}$Ni foil with a thickness of 10~$\mu$m was also irradiated in order to study the performance and properties of that material under irradiation from the WNR neutron beam. No actual targets were used for $^{nat}$Al and $^{63}$Cu as these two materials were only used in calculations to benchmark the performance of our simulation against a previously validated one based on MCNP~6.2.

The choice of the measurements utilized in this work was limited by the experimental studies performed with the LENZ instrument in only one run cycle and for which data were available. The choice of utilizing a single run cycle we made in order to ensure uniformity of the instrumental and facility-wise conditions. For instance, the position of the LENZ chamber at every run cycle determines the flight path length characteristic of the neutron TOF as well as the choice and specific geometrical setup of the detectors inside the chamber. 

\begin{table*}[t]
    \centering
    \begin{tabular}{|c|c|c||c|c||c|c|}
    \hline
    {}& \multicolumn{2}{c||}{Target Material} & \multicolumn{2}{c||}{Reactions of interest} & \multicolumn{2}{c|}{Method of study} \\
    \hline
    $\#$ & Material & Thickness ($\mu m$) & Reaction &  Q value g.s. (MeV) & Experiment & Simulation \\ \hline

    \multirow{3}{1em}{1} & \multirow{3}{3em}{Mylar} & \multirow{3}{1.5em}{15} & $^{1}$H(n,el) & 0 & \multirow{3}{2em}{YES} & \multirow{3}{6em}{GEANT4} \\
    
    {} & {} & {} & $^{12}$C(n,a) & -5.7 & {} & {} \\ 
    {} & {} & {} & $^{16}$O(n,a) & -2.2 & {} & {} \\ \hline
    
    \multirow{3}{1em}{2} & \multirow{3}{3.5em}{Kapton} & \multirow{3}{2em}{100} & $^{1}$H(n,el)& 0 & \multirow{3}{2em}{YES} & \multirow{3}{6em}{GEANT4}\\
    
    {} & {} & {} & $^{12}$C(n,a) & -5.7 & {} & {} \\
    {} & {} & {} & $^{14}$N(n,p) & 0.62 & {} & {} \\ \hline
    
    \multirow{2}{1em}{3} & \multirow{2}{3em}{$^{nat}$Ni} & \multirow{2}{1.5em}{10} & $^{58}$Ni(n,p) & 2.9 & \multirow{2}{2em}{YES} & \multirow{2}{6em}{GEANT4 $\&$ MCNP6.2}\\
    
    {} & {} & {} & $^{58}$Ni(n,a) & 0.4 & {} & {} \\ \hline
    
    \multirow{2}{1em}{4} & \multirow{2}{2.5em}{$^{63}$Cu} &  \multirow{2}{1em}{1} & $^{63}$Cu(n,p) & -0.72 & \multirow{2}{1.5em}{NO} & \multirow{2}{6em}{GEANT4 $\&$ MCNP6.2} \\
    
    {} & {} &  {}  & $^{63}$Cu(n,a) & 1.72 & {} & {} \\ \hline
    
    \multirow{2}{1em}{5} & \multirow{2}{2.5em}{$^{nat}$Al} & \multirow{2}{2em}{100} & \multirow{2}{4em}{$^{27}$Al(n,p)} & \multirow{2}{2.5em}{-1.83} & \multirow{2}{1.5em}{NO} & \multirow{2}{6em}{GEANT4 $\&$ MCNP6.2} \\
    
    {} & {} &  {}  & {} & {} & {} & {} \\ 
    
    \hline
    \end{tabular}
    \caption{The target materials along with the corresponding reactions studied in this work}
    \label{tab:list_reaction}
\end{table*}




\section{The GEANT4-based simulation model.}
\label{sec:Geant}

Monte Carlo simulations are a fundamental research tool for the design and optimization of experimental configurations in their entirety. In this work we used the version 10.6.p01 of the GEANT4 toolkit~\cite{AGOSTINELLI2003250, ALLISON2016186, 1610988} to build a detailed simulation model of the LENZ experimental setup. Specifically, the simulation includes the geometry of the experimental apparatus shown in Figure~\ref{LENZ}, the LANSCE neutron beam characteristics, the interaction between the beam and the LENZ target, and the response of the detection system. The emission of secondary protons and alpha particles at neutron energies below 20~MeV was an area of specific focus. To this extent, we implemented the detailed partial cross-sections based on the ENDF/B-VIII.0 evaluated library (see Section \ref{sec:Geant_physics}). Due to the high precision we decided to include in the simulation model, it was deemed necessary to employ a probability biasing technique to reduce the simulation computing time to a practically reasonable amount. In the following subsections we describe the various components of this simulation model. 



\subsection{Experimental Geometry}
\label{sec:Geometry_description}

\begin{figure}[h]
\centering
\includegraphics[width=\linewidth]{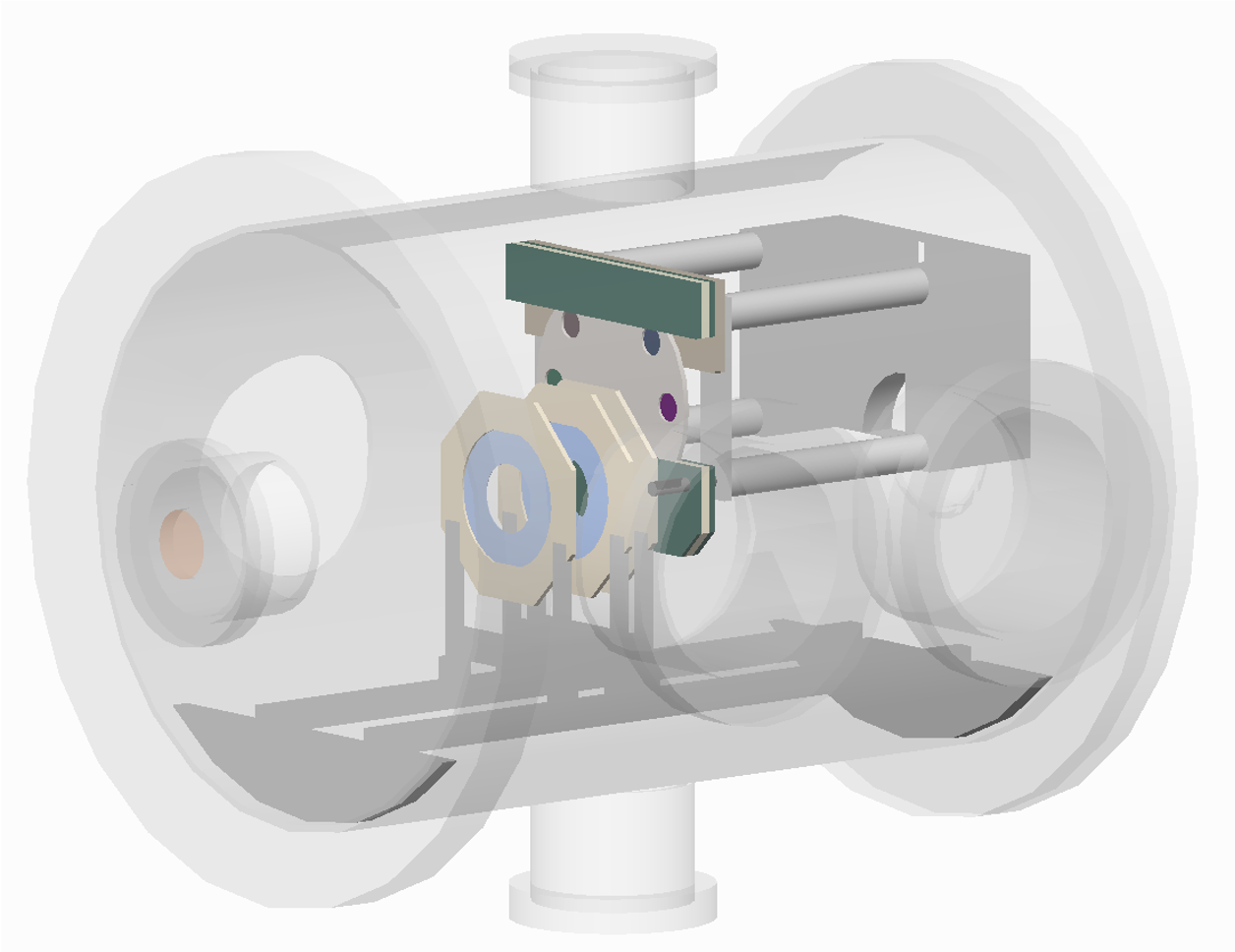}
\caption{Visualization of the LENZ chamber as made with GEANT4 simulation. The configuration of the inner chamber consists of two Kapton windows at the entrance and exit of the neutron beam, the 8 position target wheel and three annular double-sided silicon strip detectors (DSSD). The DSSDs are placed in telescope configuration, acting as $\Delta$E-E, and provide large solid angle coverage with low detection threshold. The LENZ chamber has multi-target capability and a good timing resolution, optimized for time of flight measurements.}
\label{GeantLENZ}
\end{figure}

The geometry of the LENZ experimental setup as implemented in the simulation code is presented in Figure \ref{GeantLENZ}. Every part of the chamber was reproduced in the simulation based on the engineering designs shown in Figure \ref{LENZ}, including the beam-line stand that supports the LENZ chamber. While surrounding structures (such as the chamber stand) are not crucial for the simulation of the charged-particle detection experiments, they can potentially contribute to the neutron-induced background and hence were included in the simulated geometry. On the other hand, the signal cables, pre-amplifiers and couplings to the vacuum pump and gauges were considered less significant sources of neutron-induced background and were not included in the geometry.

The configuration of the interior of the chamber in Figure \ref{GeantLENZ}, reproduces the experimental setup used in the run cycle of 2017, when the samples examined in this work were taken. Perpendicularly to the beam axis and in forward angles from the target, three annular double-sided silicon strip detectors (DSSDs) were set up for that run cycle. The DSSDs were placed at distances of 2.0~cm, 4.2~cm and 9.2~cm downstream of the target, with increasing thicknesses of 65~$\mu$m, 300~$\mu$m and 500~$\mu$m respectively. These detectors covered a wide range of forward angles, between 14$^{\circ}$ and 67$^{\circ}$ degrees. In this run cycle, 76~$\mu$m thick Kapton foils were used as chamber windows at the entrance and exit of the neutron beam.

\subsection{Neutron Beam}
\label{sec:Neutron_beam_geant}

As mentioned before, the LENZ chamber is placed at flight-path 15R, which is extending downstream of the spallation target at an angle of 15$^{\circ}$ degrees to the right with respect to the direction of the proton beam. The energy dependent flux of neutrons at the entrance of the LENZ chamber was obtained from the result of the MCNP simulation, described in Section~\ref{sec:Neutron_beam_LANSCE}. The neutron energy vs. flux histogram, shown in Figure~\ref{Beam}, was sampled in GEANT4 to generate primary neutron energies. The neutron beam profile is implemented as a pencil beam 2.0~cm in diameter with a zero degree incident angle and zero angular divergence, to approximate the neutron beam shape, after passing through the various beam-line collimators in the 15R flight-path of WNR and while arriving at the entrance of the LENZ chamber. 
 

\subsection{Detector responses}
\label{sec:Detector_responses}

The LENZ instrument includes a number of annular double-sided silicon strip detectors~\cite{Micron} that can be placed on beam axis upstream and downstream of the target. Three such detectors have been implemented in the simulation and are shown in Figure \ref{GeantLENZ}. Each of the detectors has a disc-shaped active area with an outer diameter of 96~mm, and an inner diameter of 48~mm. Their inner hollow part makes them ``blind" to any incoming particle (neutron or charged particle) at small forward angles. Moreover, the energy and time resolution along with the dead layer, which varied from 100~nm to 300~nm of Al layer depending on the detector's thickness, have been included in the simulation. 

The energy resolution of the DSSDs was determined in a previous work \cite{LeePSD2019} by using a $^{229}$Th source, which emits alphas from 4 MeV to 9 MeV. The detector resolution was found to vary quadratically as a function of energy and, therefore, a second order polynomial was implemented to reproduce the energy dependence of the detector resolution in the simulation. 


The timing resolution of the DSSDs for the detected charged particles is generally determined by processing the characteristic waveforms from the observed gamma flash, generated from the WNR spallation neutron production target. More specifically, the spallation source produces prompt $\gamma$-rays which after traveling through the flight path can be detected in LENZ as scattered gammas. In the signal post-processing by applying appropriate optimized digital filters on the signal waveform, we can deduce a triggered time stamp and a pulse height. The intrinsic detector timing resolution that is deduced in this way for each incoming pulse height typically has a slight energy dependence. As far as the simulation is concerned, however, this slight energy dependence was considered negligible and the timing resolution implemented as a Gaussian broadening with a width corresponding to the Full Width at Half Maximum (FWHM) of 4~ns obtained from the experimental data of \cite{LeePSD2019} was, for all practical purposes, adequate for a precise description of LENZ spectra.

\subsection{Physics implementation}
\label{sec:Geant_physics}

All of the physics processes and data were implemented to the simulation through the reference physics list \textit{FTFP\_BERT\_HP\_EMZ}. Electromagnetic processes such as electronic energy loss of ions were implemented by using the \textit{EmStandardPhysics\_option4}, which is an optimized electromagnetic physics list for low energy applications that require high accuracy on hadron and ion tracking. For more accurate stopping power calculations of protons and alpha particles at low energies, according to the suggestions of references \cite{Ivanchenko2017, VAGENA202044}, the default step limit function was lowered to 0.01~$\mu$m. 



Neutron interactions with various materials were included using the GEANT4 High Precision Physics libraries. In order to have a standardized and realistic modeling of neutron-induced reaction yields we made use of a modified nuclear data evaluation file as an input to the simulation, built upon reference total and partial differential cross sections derived from the ENDF/B-VIII.0 nuclear data library, as discussed in~\cite{KIM2020163699}. This modified evaluation file includes an improved treatment of the outgoing charged particle energy-angle spectra from (n,p) and (n,$\alpha$) reactions for more than 60 nuclei including Ni isotopes of interest for this work. Besides these 60 nuclei, for all other isotopes we used the inputs from the ENDF/B-VIII.0 evaluation that were already  available for use with GEANT4~\cite{ENDFVIII_2018, ENDFVIII_2012, Mendoza20142357}.

During the process of developing the simulation, some limitations in the treatment of the nuclear data inputs in GEANT4 had to be overcome. One was related to the handling of empty library records that occur for reactions which are not included in the evaluated files. Empty records exist for certain types of reaction, or neutron beam energies. For example, the main beam energy related limitation, is for neutron-induced reactions above 20~MeV for which ENDF/B-VIII.0 often does not contain evaluated cross section data. This fact limited the high accuracy of our simulated results above 20~MeV, since when exceeding this energy, reaction yields are handled by an approximate hadronic model~\cite{HadronModels, Bertini}. Consequently we have limited our discussion to energies below 20~MeV. 

The second major issue was encountered when the outgoing energies in the simulated spectra did not correspond to the expected energies based on reaction kinematics, even after applying the modified evaluations that had previously been validated using an MCNP simulation of the LENZ setup. The culprit appeared to originate from within the GEANT4 source code itself and the routine that handles the kinematics from the reactions of interest when partial differential cross sections are provided (given as MF = 4 in the ENDF format). The file under question, G4ParticleHPInelasticCompFS.cc, handles both neutron inelastic scattering as well as reactions such as (n,p) and (n,$\alpha$). The routines included in this file have been frequently revised in recent versions/patches of GEANT4, however, as of GEANT4.10.6, issues still remain. The most important change that we had to make was to correctly sample the reaction Q-value since the routine was still assuming that it was equal to zero minus the excitation energy of the populated state. This is only valid if the ground state Q-value is zero, as is the case for elastic scattering. Hence, for reactions with positive ground state Q-values like $^{58}$Ni(n,p)$^{58}$Co, the outgoing energies were all shifted down in energy in the default GEANT kinematics calculation (see discussion in section \ref{sec:Comparison backing}). Also, we found that the population of the ground states were being suppressed entirely because the code was only treating the excited levels. This is most likely attributed to the fact that the same routine also treats inelastic neutron scattering, while elastic scattering to the ground state is treated in a different part of the code. 

\subsection{Cross Section Biasing - Implementation and Limitations}
\label{sec:Cross_section_biasing}

In order to render the statistics in the simulated results comparable to the ones of the experimental spectra in a reasonable amount of computing time, all the corresponding reaction cross-sections had to be enhanced via the predefined generic biasing technique provided by the GEANT4 toolkit. The technique takes advantage of the approximate relationship that connects the mean interaction length with the so-called analog cross-section,
\begin{equation}
    \sigma_{analog}= \frac{1}{\lambda(E)} \label{eq:AnXS}
\end{equation}
to enhance the cross section by altering the interaction length and hence to increase the occurrence of the individual physics process interactions.
In this work, the cross section enhancement was applied by biasing the individual physics process interaction occurrence \cite{ALLISON2016186}. This generic biasing scheme is optimized for neutral particles, which have no continuous energy loss along a step. 
 
The interaction length is defined as the length which the beam covers within a defined volume of material (importance volume) before a reaction occurs, and can be expressed in terms of the total cross section as:
\begin{equation}
\centering
  \lambda(E) = \left(\sum_{i}{[n_i\sigma(Z_i,E)]}\right)^{-1} 
  \label{eq:intLen}
\end{equation}
where $\sigma(Z,E)$ is the total cross section of the process per atom, $\sum_{i}$ sums over all elements in the material of the importance volume, and $n_i$ is the number of atoms per volume for the i$^{th}$ element in a compound material and is calculated as:
\begin{equation}
\centering
  n_i = \frac{\mathcal{N} \rho w_i}{A_i} \label{eq:density}
\end{equation}
where $\mathcal{N}$ is Avogadro's number, $\rho$ is the density of the medium, $w_i$ is the proportion by mass of the $i^{th}$ element and $A_i$ is  the molar mass of the $i^{th}$ element.

When decreasing the interaction length by a \textit{(biasing)} factor ($F_{bias}$), the probability of a reaction to occur increases in a proportional way. 
\begin{equation*}
\sigma_{bias} = F_{bias} \times{\sigma_{analog}}
\end{equation*}
\begin{equation*}
= \frac{F_{bias}}{\lambda(E)}    
\end{equation*}
\begin{equation}
\sigma_{bias}= F_{bias}\sum_{i}{[n_i\sigma(Z_i,E)]} 
\label{eq:BiasXS}    
\end{equation}
effectively increasing the cross section $\sigma$ by the biasing factor.
The above described biasing technique can be extended to include different parts of the experimental setup, as e.g. the chamber walls, the entrance foil, etc., to account for possible low-probability neutron scattering events that increase the induced background in the detectors. The described cross section biasing technique affects the probability of any physics process to occur, including energy loss, and hence it is best suited for neutral particles like neutrons. 

However, even when using neutral projectiles, one has to be aware of the maximum reliable biasing factor that can be applied; especially when the stopping power of a reaction product is of great importance for the simulated project. The reason lies in the fact that when the biasing factor exceeds a limit, the reactions tend to occur earlier (i.e. at a shallower depth) within the importance volume. 

\begin{figure}[h!]
\centering
\includegraphics[width=\linewidth]{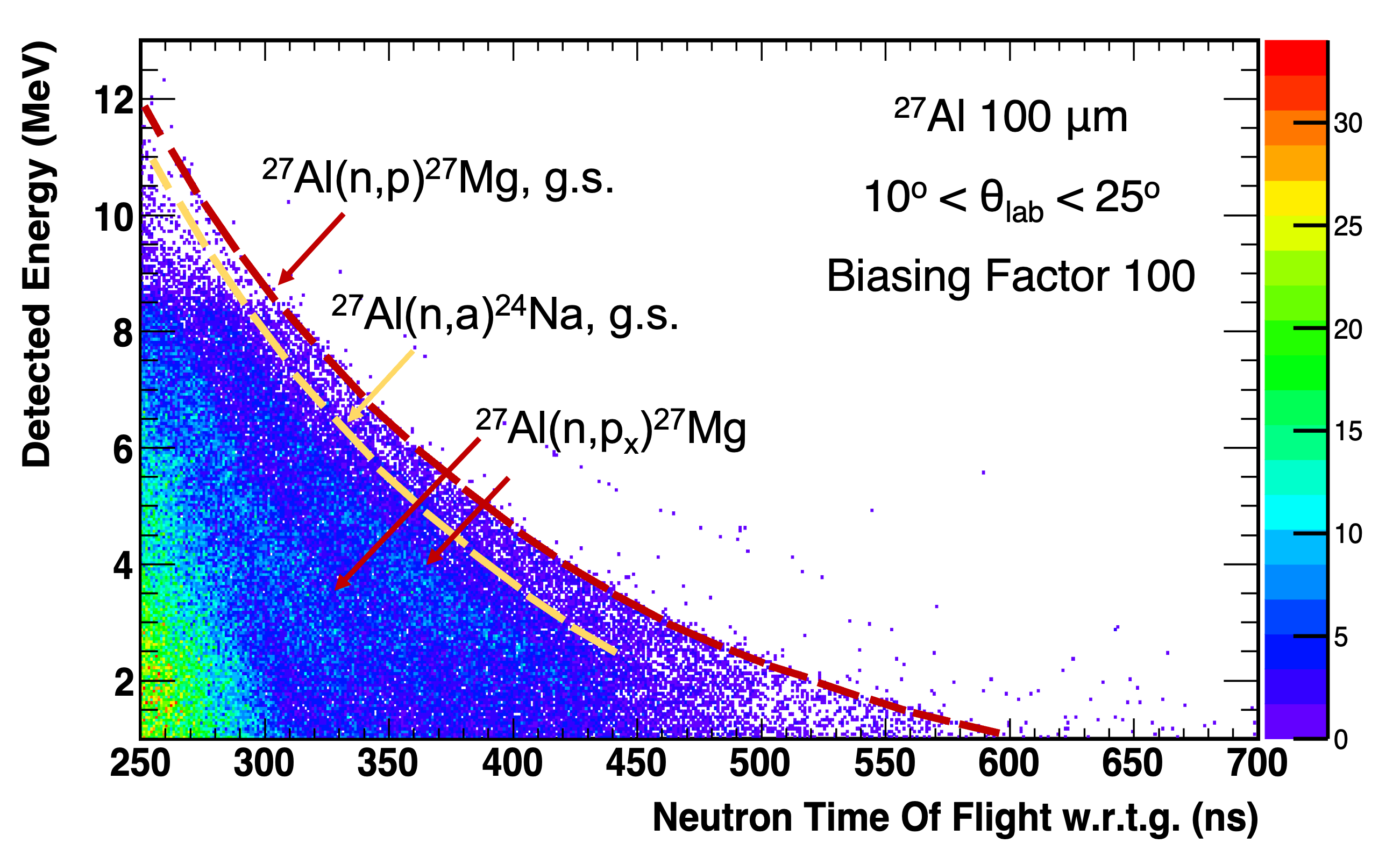}
\includegraphics[width=\linewidth]{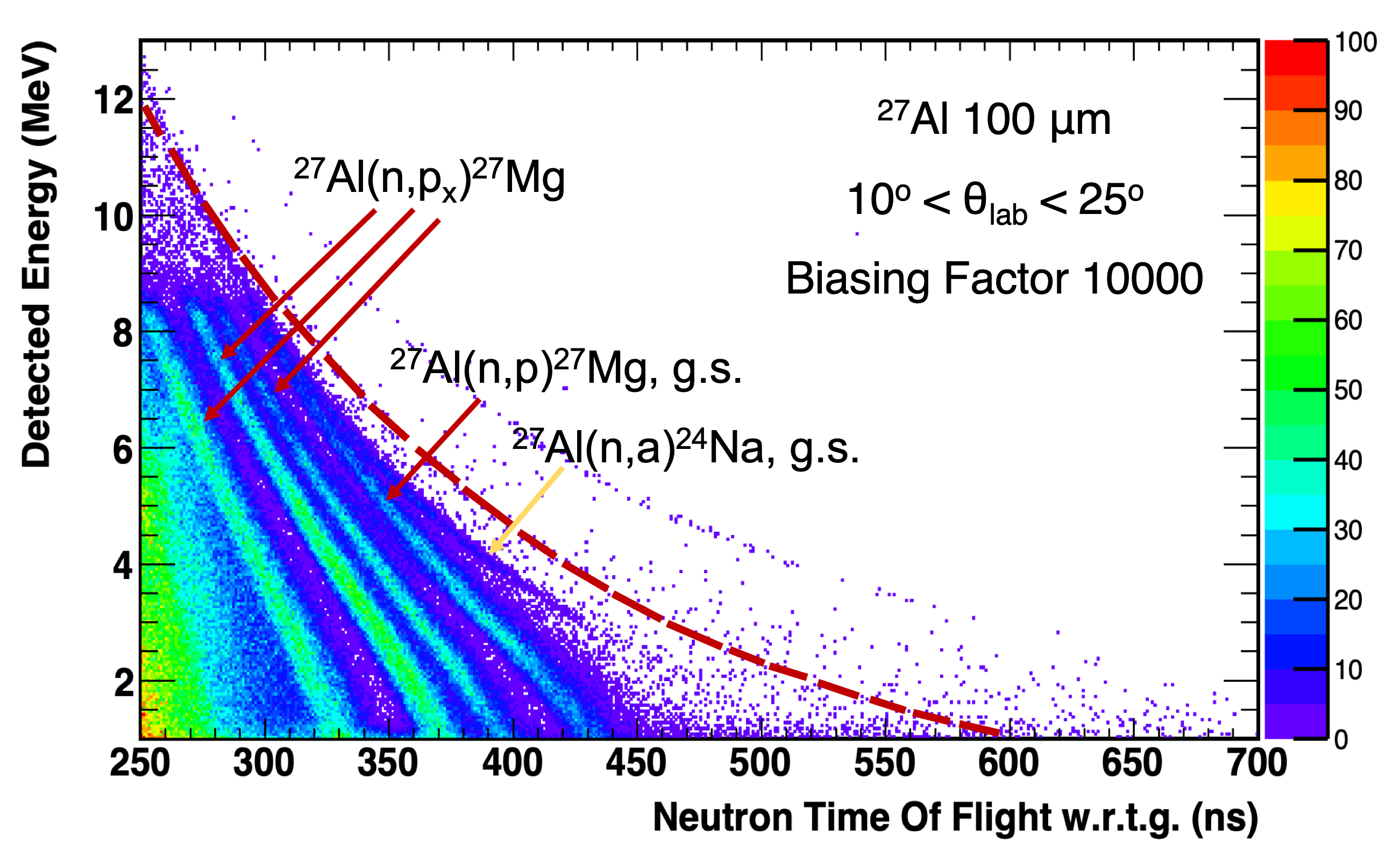}
\caption{The effect of the cross section biasing on simulated spectra for neutron induced reactions on a 100~$\mu$m thick $^{27}$Al target. (Top) In this spectra the biasing factor was kept at the maximum limit, at 100. (Bottom) For this spectra the biasing factor used was 100 times larger than the acceptable upper limit. The spectra show that, an exaggeration of the neutron cross section biasing, results in large discrepancies, since all the neutron induced reactions are forced to occur in smaller depths inside the target, and the rest of the volume then acts as an absorber.}
\label{Al_TOF}
\end{figure}

For instance, a simple example presented in Figure \ref{Al_TOF} shows how the results can get unphysically distorted when the biasing factor applied is highly overestimated. We considered a 100~$\mu$m thick aluminum target placed in the LENZ chamber and run our simulation with two different biasing factors. In the spectra the energy of the charged particles detected from a DSSD placed 9.2~cm downstream the target as a function of the neutrons' time of flight is presented accordingly. In the case of Figure \ref{Al_TOF} (Top) the biasing factor implemented was equal to 100, while for the Figure \ref{Al_TOF} (Bottom) it was 100 times greater and equal to 10000. As expected and shown on Figure \ref{Al_TOF} (Top), due to the stopping power of the 100~$\mu$m thick target, the proton bands that result from (n,p) reactions on Aluminum are not distinguishable but rather overlapping on one another and spread throughout the whole range of the neutron time of flight. This is not however, the case for the spectrum shown in Figure \ref{Al_TOF} (Bottom), which presumably should present the same effect of stopping power, considering that the target and geometry used in both cases were completely identical. The higher biasing factor applied to this simulation has apparently altered the results and it appears as if the target used was much thinner than its actual size. As it can be observed, the $^{27}$Al(n,p) reaction along with its excited states have clearly distinguishable bands, but compressed to the lower neutron time of flight - higher beam energy region. Consequently, the neutron induced reactions occur only within the first few ~$\mu$ms of material inside the target, as the cross sections for neutrons were over-biased. Therefore, the rest of the importance volume acts rather like an absorber, with the stopping power of the non homogeneously produced charged particles being seriously overestimated. 

\begin{figure*}[h!]
\centering
 \includegraphics[width=0.3\linewidth]{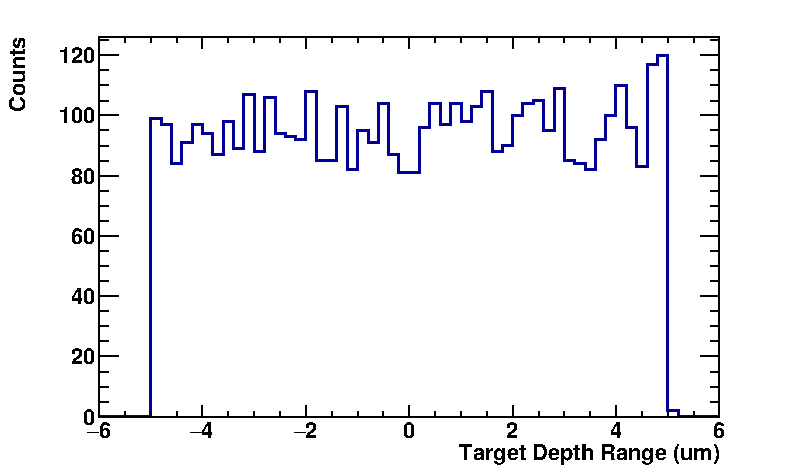}
 \includegraphics[width=0.3\linewidth]{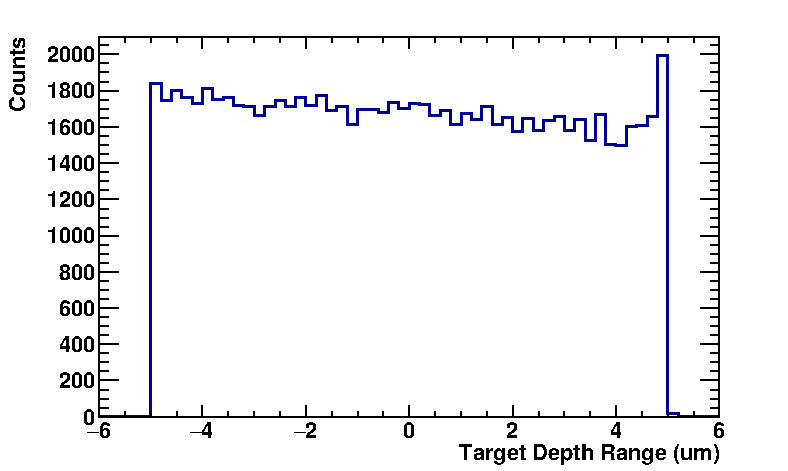}
 \includegraphics[width=0.3\linewidth]{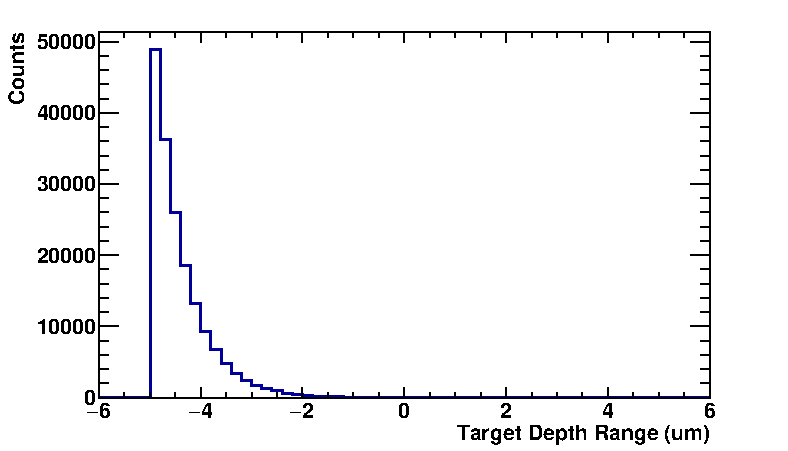}
\caption{The effect of the cross-section biasing on simulated reaction product yields as a function of target depth for 5~MeV neutrons impinging on a 10~$\mu$m (6.0E+19 thick $^{27}$Al target. In plot (a) there wasn't any biasing applied to the simulation. For plot (b) the biasing factor was set at 1200, while in plot (c) the biasing factor used was set at 100 times higher, namely at 120000. The spectra show that an exaggeration of the cross-section biasing, yields large discrepancies, since practically all the neutron-induced reactions are forced to occur at shallower depths inside the target, and the rest of the target volume then acts as a mere absorber. }
\label{BiasingTestAlResults}
\end{figure*}

Figure~\ref{BiasingTestAlResults} shows an example of the effect of different biasing factors on the reactions distribution within the target volume. For the first plot, Figure~\ref{BiasingTestAlResults}(a), no biasing factor was applied, so the distribution of the reactions is uniform, though the statistics of the plot are significantly lower than the ones implementing the biasing technique. In the second plot of Figure~\ref{BiasingTestAlResults}(b) a biasing factor of 1200 is applied and it appears that the uniformity of the distribution of the reactions is not significantly disturbed. However, in the third plot, Figure~\ref{BiasingTestAlResults}(c), with a biasing factor of 120000, the distribution is obviously distorted with the majority of reactions occurring at a smaller beam penetration depth inside the target.

For the simulations in this work the maximum acceptable biasing factor was determined by calculating the spatial distribution of reaction occurrences within targets of varying composition and thickness using increasing biasing factors. The criterion for an acceptable biasing factor was that the mean of the spatial distribution of events agrees within 10\% with the mean of the unbiased calculation. 
Although the calculations for this test were performed for mono-isotopic targets and using a mono-energetic neutron beam with an energy of 5~MeV, the results apply to more complicated target materials and broader neutron beam energy range.

\begin{figure}[h!]
\centering
\includegraphics[width=\linewidth]{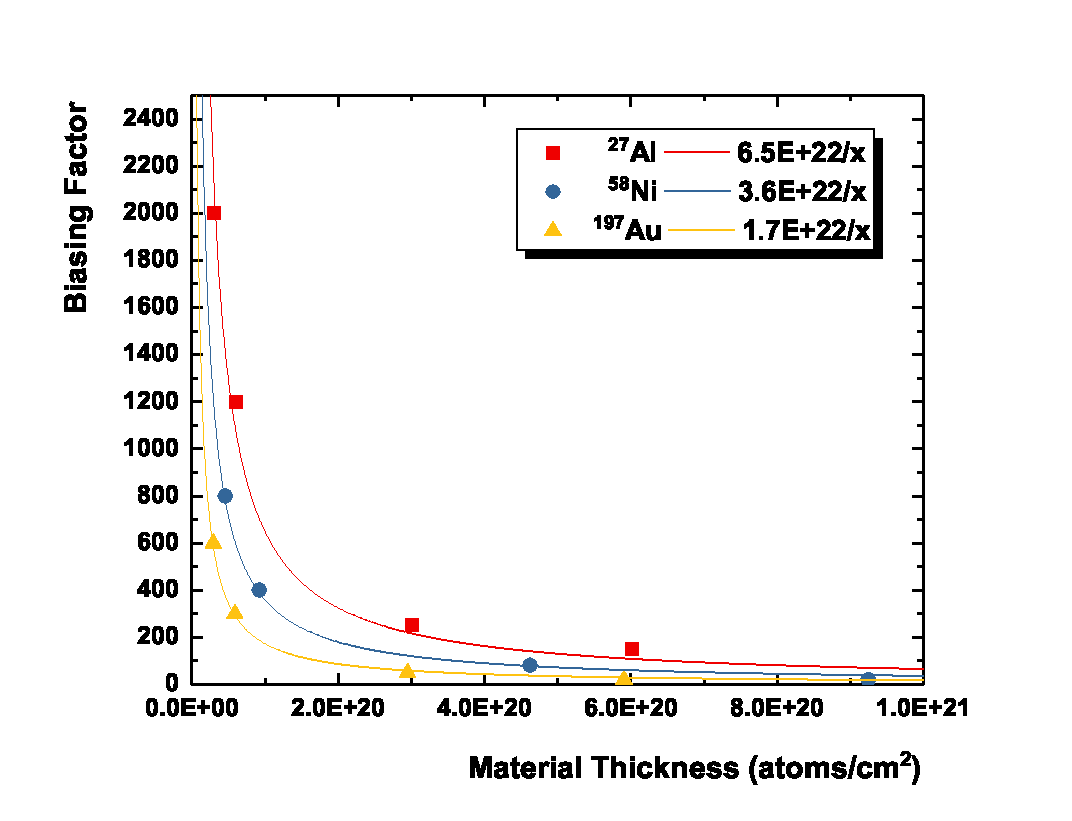}
\caption{The biasing factor as function of thickness expressed in at/cm2 for three different target isotopes. The solid lines correspond to various fits of the data with a $f(x)=a/x$ fit function. The upper acceptable limit of the biasing factor was determined by calculating the spatial distribution of reaction occurrences within targets of varying composition and thickness (see text for details).}
\label{BiasingTestLimits}
\end{figure}

Figure~\ref{BiasingTestLimits}, shows the dependence of the maximum acceptable biasing factor on the thickness of the target (expressed as an areal density atoms/cm${^2}$) for three different elements. The colored lines represent fits of the maximum acceptable biasing factor values versus the material thickness. The fit in all cases is consistent with the maximum biasing factor varying in an inversely proportional way to the material thickness. The constant factor depends on experimental parameters such as the neutron energy and seems to roughly scale with the inverse of the target density.  

\section{Results and Discussion} 
\label{sec:Results and Discussion}

To benchmark our newly developed GEANT4 application we took advantage of a pre-existing, validated MCNP-based simulation of LENZ, comparing the outcome of both codes when using a $^{63}$Cu reference target. To confirm the reliability of the simulated response, we recreated sample cases based on a previous run cycle of LENZ and then compared them against experimental data. In this section we present our study, on the qualitative comparison of the simulated yields with the experimental ones, obtained using two standard backing materials for nuclear physics experiments, such as Kapton and Mylar. In addition, we present our attempt of simulating a target sample with many different stable isotopes, such as $^{nat}$Ni. The detector configuration used to obtain all the aforementioned data was as described in section \ref{sec:Geometry_description}.

\subsection{Benchmark of the GEANT4 results for LENZ using MCNP}
\label{subsec:Benchmark}

\begin{figure}[h!]
\centering
\includegraphics[width=\linewidth]{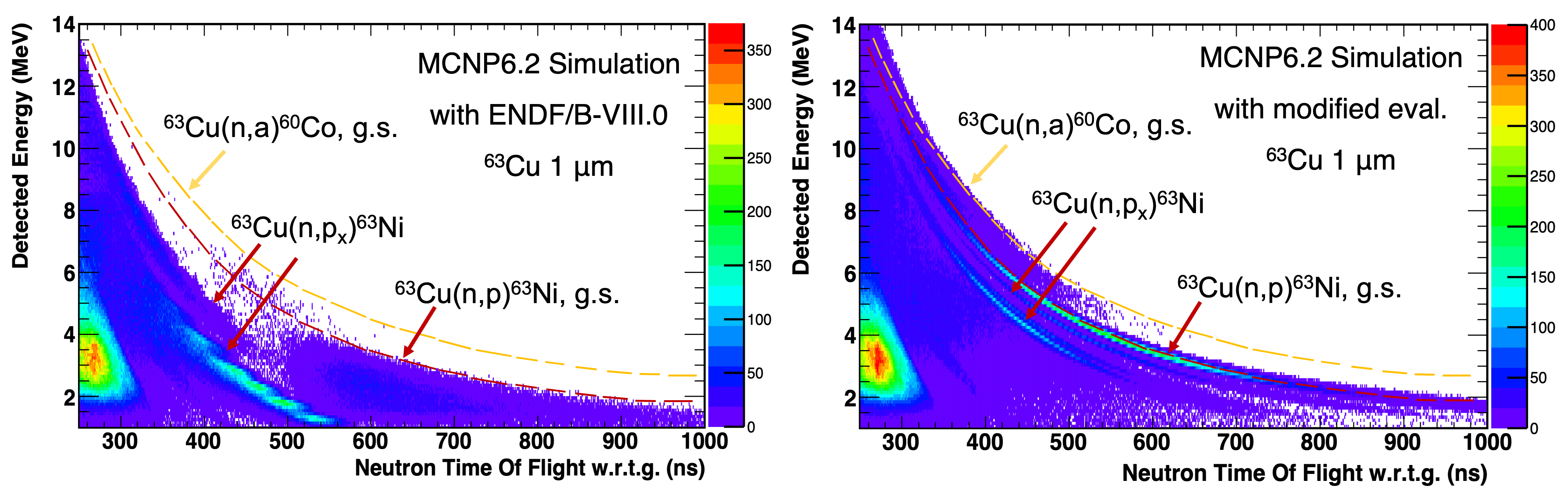}
\includegraphics[width=\linewidth]{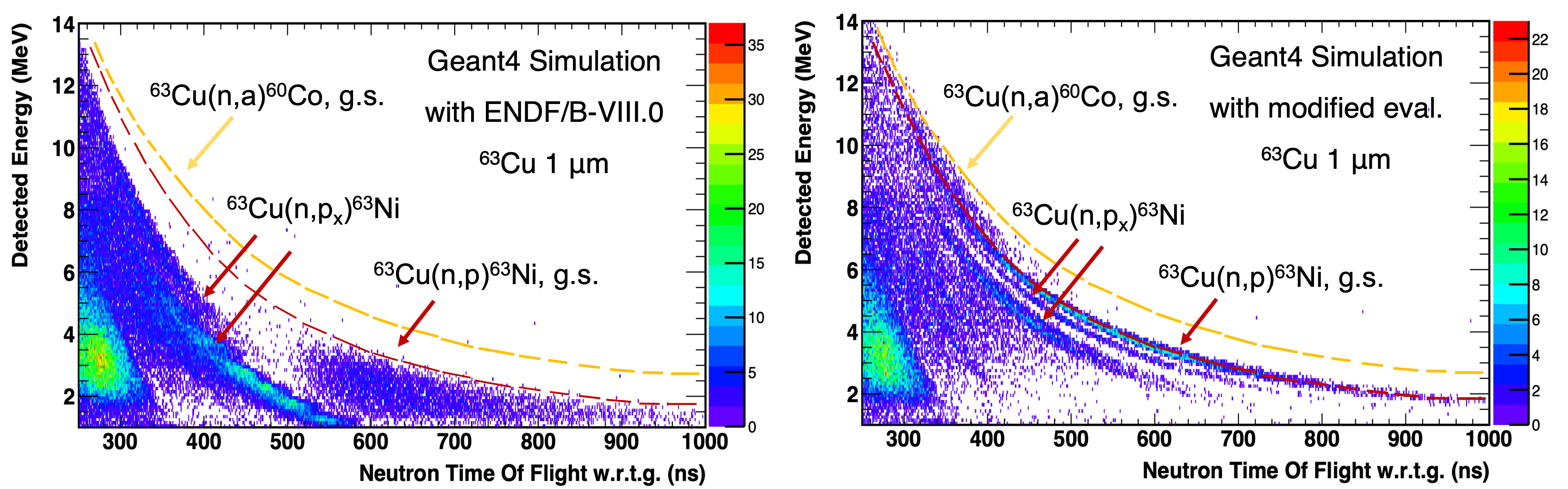}
\caption{Charged particle energy deposition in the detector at 9.2~cm as a function of incident neutron time of flight, from a mono isotopic $^{63}$Cu target of 1~$\mu$m thickness as simulated in MCNP6.2 (Top) and GEANT4 (Bottom).}
\label{63Cu_Geant}
\end{figure}

The spectra presented in Figure~\ref{63Cu_Geant}, show the simulated results of the $^{63}$Cu target with a thickness of 1~$\mu$m as obtained using the GEANT4 and MCNP6.2 simulations of LENZ respectively, and utilizing two different nuclear data libraries as inputs. For this case, ENDF/B-VIII.0 models the outgoing charged particle spectra from (n,p) and (n,$\alpha$) using double differential energy-angle (DDX) spectra whereas the modified evaluation includes partial differential cross sections, with respect to angle, for the population of the discrete levels (up to the 40$^{th}$) in the residual nuclei. The remainder of the levels are combined to form the ``continuum" contribution, for which DDX information provides an average description of the outgoing charged particle energies. The labelled kinematic curves show the expected outgoing charged particle energy as a function of neutron time of flight for populating the ground states in $^{60}$Co and $^{63}$Ni and assuming that the reaction occurred at the downstream face of the target. Here, the modified evaluation provides a better description of the expected outgoing energies as validated by past LENZ experiments with a brass target \cite{KIM2020163699}. 

\subsection{Comparison of LENZ simulation with experimental data}
\label{sec:Comparison backing}
The results of the benchmarking exercise show that the two simulations treat the very different nuclear data inputs in a consistent way. We are thus enabled to proceed and compare the GEANT4 simulations with experimental data taken with the LENZ detector. The comparison focused on the simulation of common backing materials and of a representative multi-isotopic target. The backing materials used in nuclear physics experiments for depositing targets of interest are usually thin foils or films. For the purposes of this study, we chose two carbon and hydrogen containing foils, Mylar, and Kapton, with very different thickness each, to test the accuracy of the simulation results for the neutron beam and target interactions of relevance that typically appear during nuclear physics experiments. 


\begin{figure}[h!]
\centering
\includegraphics[width=\linewidth]{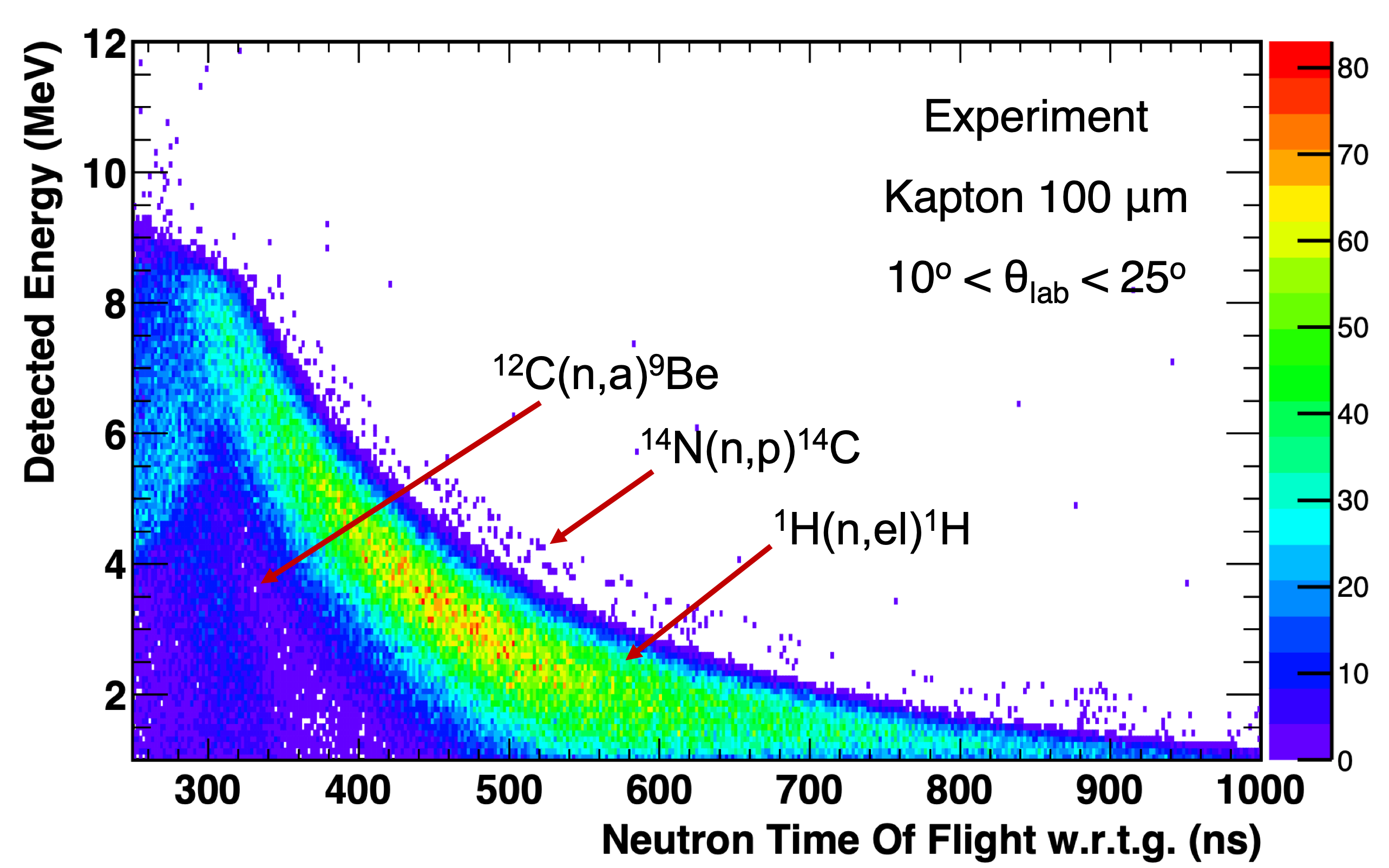}
\includegraphics[width=\linewidth]{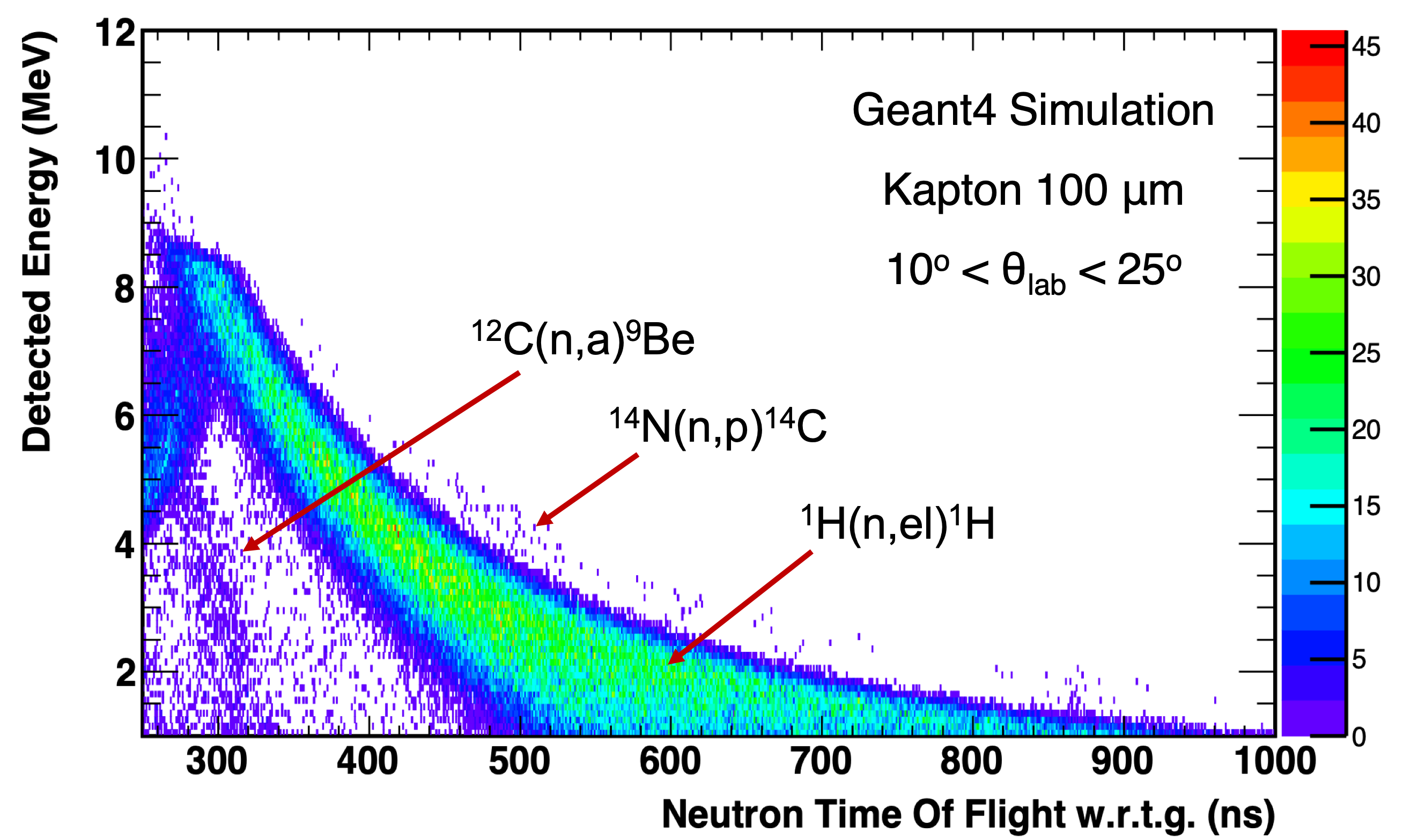}
\caption{Charged particle energy deposition as a function of incident neutron time of flight, in the detector at 9.2~cm downstream of a 100~$\mu$m thick Kapton foil for the (Top) Experimental data and (Bottom) a GEANT4 simulation implementing a biasing factor of 50. The main characteristic in both spectra is the broad band of the forward scattered proton recoils that are produced by the Kapton foil, and distributed throughout the whole range of neutron time of flight. Contributions from neutron-induced reactions with nitrogen and carbon in the Kapton foil are also displayed and labeled accordingly.}
\label{Kapton}
\end{figure}

The experimental results obtained by the irradiation of the Kapton foil are presented at the top pane of Figure~\ref{Kapton}, while the simulated ones are shown at the bottom of the same figure. The spectra show the energy of the charged particles that was detected by the DSSD detector that was placed at 9.2~cm away from the target, as a function of the neutron time of flight. The main characteristic in both experimental and simulated spectra is the broad band of proton recoils (labeled $^1$H(n,el)$^1$H), distributed throughout the whole range of neutron time of flight. This band can be mainly attributed to forward scattered protons from the elastic scattering of neutrons in the Kapton target. The simulation can recreate the width and the position of the proton band in agreement with the experimental data, indicating that the stopping power, energy, and lateral straggling inside the target are properly reproduced in the simulation code. 

A characteristic feature of the experimental data which is, at least to a comparable extent, absent from the simulated spectra is the amount of background counts at low (below 400~ns) neutron time of flight values. This well-known source of background events present in all the experimental spectra is attributed to  neutrons that are scattered on flight path elements which define the neutron beam shape, such as the cleanup collimator, the steel shutter system, etc. These parts of the neutron beam flight path, it was not practically possible to model in our GEANT4 application, and hence their effect is not included in the simulated spectra.

\begin{figure}[h!]
\centering
\includegraphics[width=\linewidth]{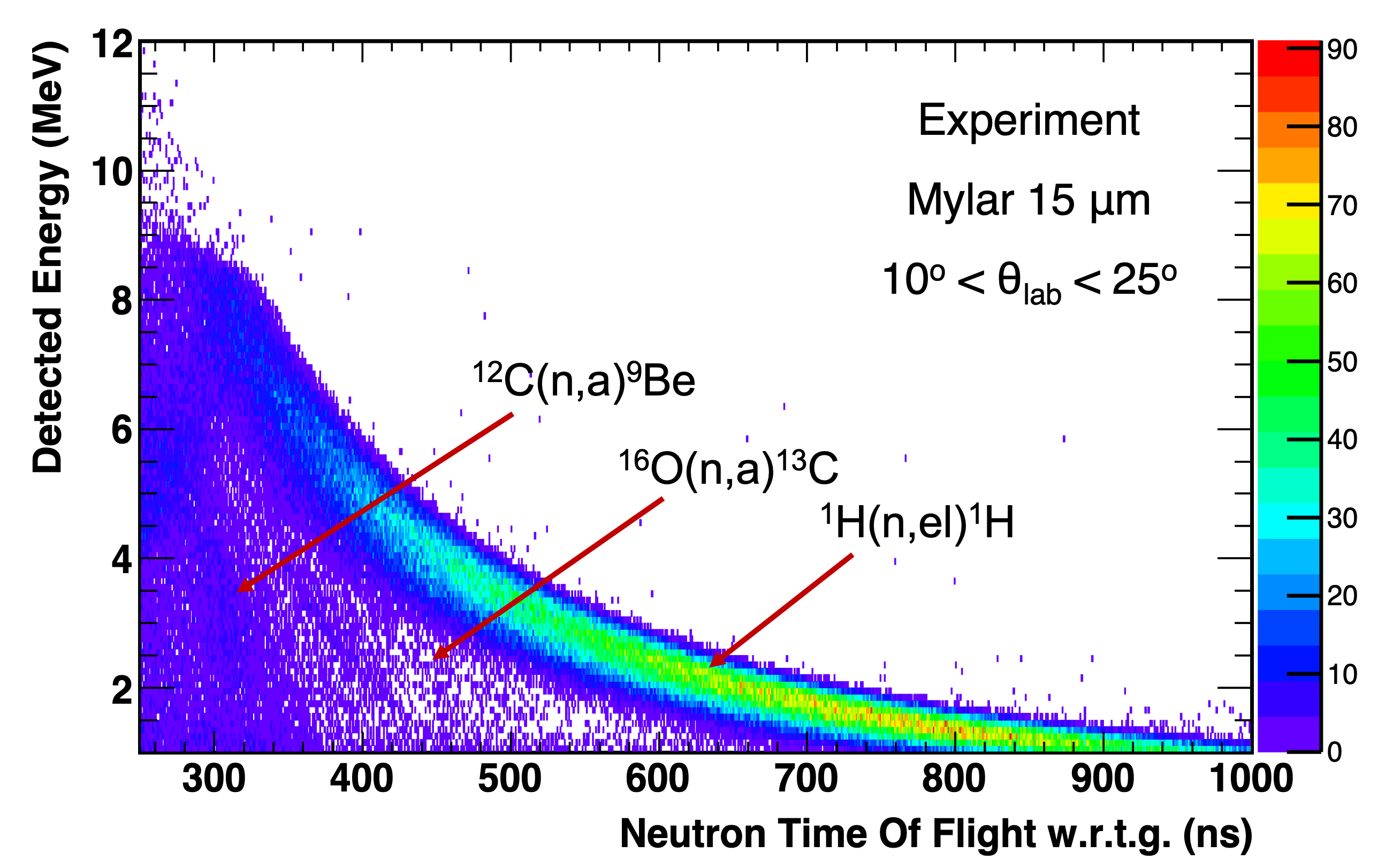}
\includegraphics[width=\linewidth]{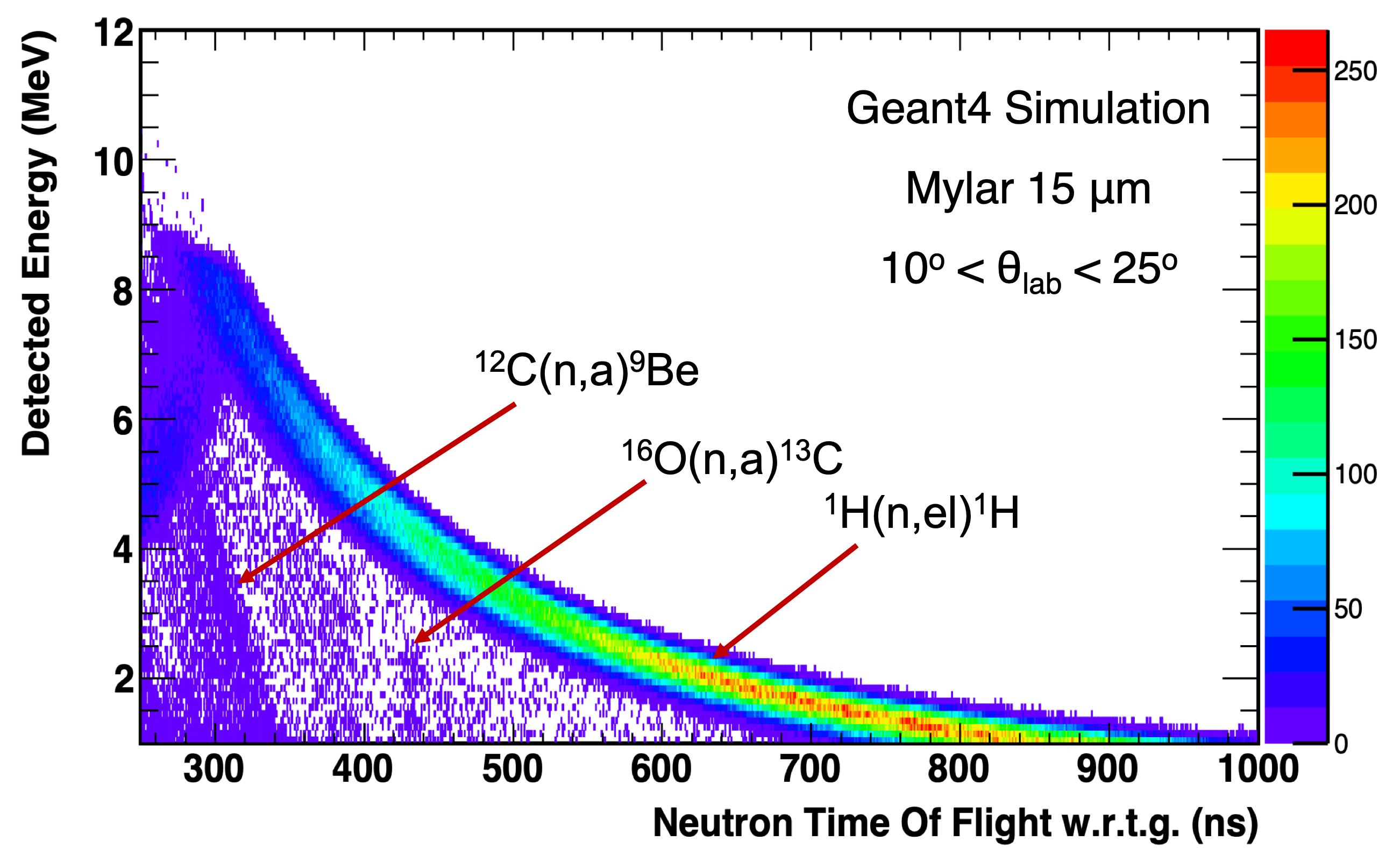}
\caption{Top: Experimental spectrum of the energy deposited by charged particles as a function of incident neutron time of flight in the detector placed 9.2~cm downstream of the Mylar foil. Bottom: GEANT4 simulation of the same spectrum (implementing a biasing factor of 500). The simulation is in fair agreement with the experimental data. Both spectra are dominated by the broad band of forward scattered proton recoils caused by the $^1$H(n,el)$^1$H reaction on the hydrogen content of the Mylar foil. These events are distributed throughout the whole range of neutron time of flight. Contributions from neutron-induced reactions with oxygen and carbon in the Mylar foil are also displayed and labeled accordingly.}
\label{Mylar}
\end{figure}

The results for the Mylar foil as obtained from the detector positioned furthest from the target (9.2~cm) are shown in Figure \ref{Mylar} (Top), while the results from the GEANT4 simulation in Figure \ref{Mylar} (Bottom). 

The spectra are in a quite satisfactory agreement. In the case of Mylar, apart from the flight path-induced background events, there was an additional background contribution, which was due to the Kapton tape used to mount the Mylar foil on the target frame. Part of the tape intercepted the neutron beam path, yielding more scattered protons, which have subsequently partially contaminated the spectrum.

Nonetheless, a closer inspection of both spectra reveals that the bands of the forward elastic scattered protons, showing a peak at around 380~ns, appear exactly at the same position and present the same distribution. Another distinct characteristic of the spectra is the existence of punch through protons (protons that have partially lost energy in the detector) and which appear from 380~ns and extend all the way down to 250~ns, indicating that the proton stopping power inside the detector is well reproduced. The alpha particles corresponding to the events shown below the proton band in the spectra, originate from (n,a) reactions mainly on carbon and to a lesser extent on oxygen, and have generally low statistics in both spectra. The alpha particles created from the $^{12}$C(n,a) reaction can be clearly spotted in the simulated spectrum, as their band lies below the proton one, starting at around 340~ns and extending up to 400~ns, implying deposited energies up to 13~MeV. However, the same events are not clearly distinguishable in the experimental spectrum due to the background events at the low neutron time of flight region mentioned above.

It should be noted that in the case of Mylar the adequate neutron-proton forward elastic scattering adopted by the corresponding GEANT4 hadron physics list is the main observable. The angular distribution for the ground state of the $^{16}$O(n,a) reaction was obtained from the ENDF/B-VIII.0 evaluated library, although this is not the case for the $^{12}$C(n,a) one, as there do not exist any available angular distribution data. Therefore, the $^{12}$C(n,a) reaction is considered to have an isotropic angular distribution in the simulation. In any case, the few resulting alphas from these reactions, are blended with the abundant protons produced from the elastic scattering and are not easily distinguishable, especially in the experimental spectra, rendering their respective detailed study quite difficult. 

\begin{figure}[htp]
\centering
\includegraphics[width=\linewidth]{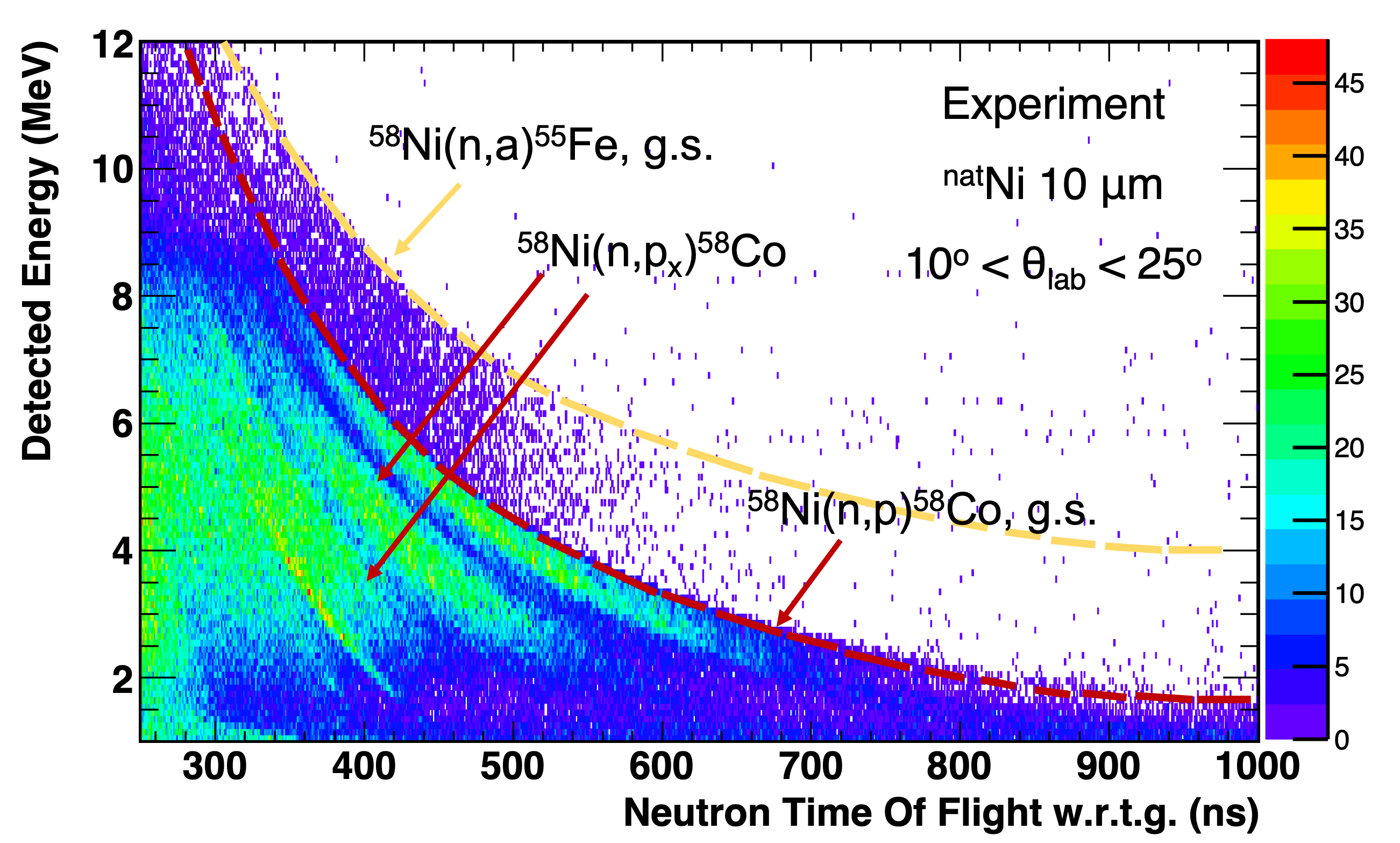}
\includegraphics[width=\linewidth]{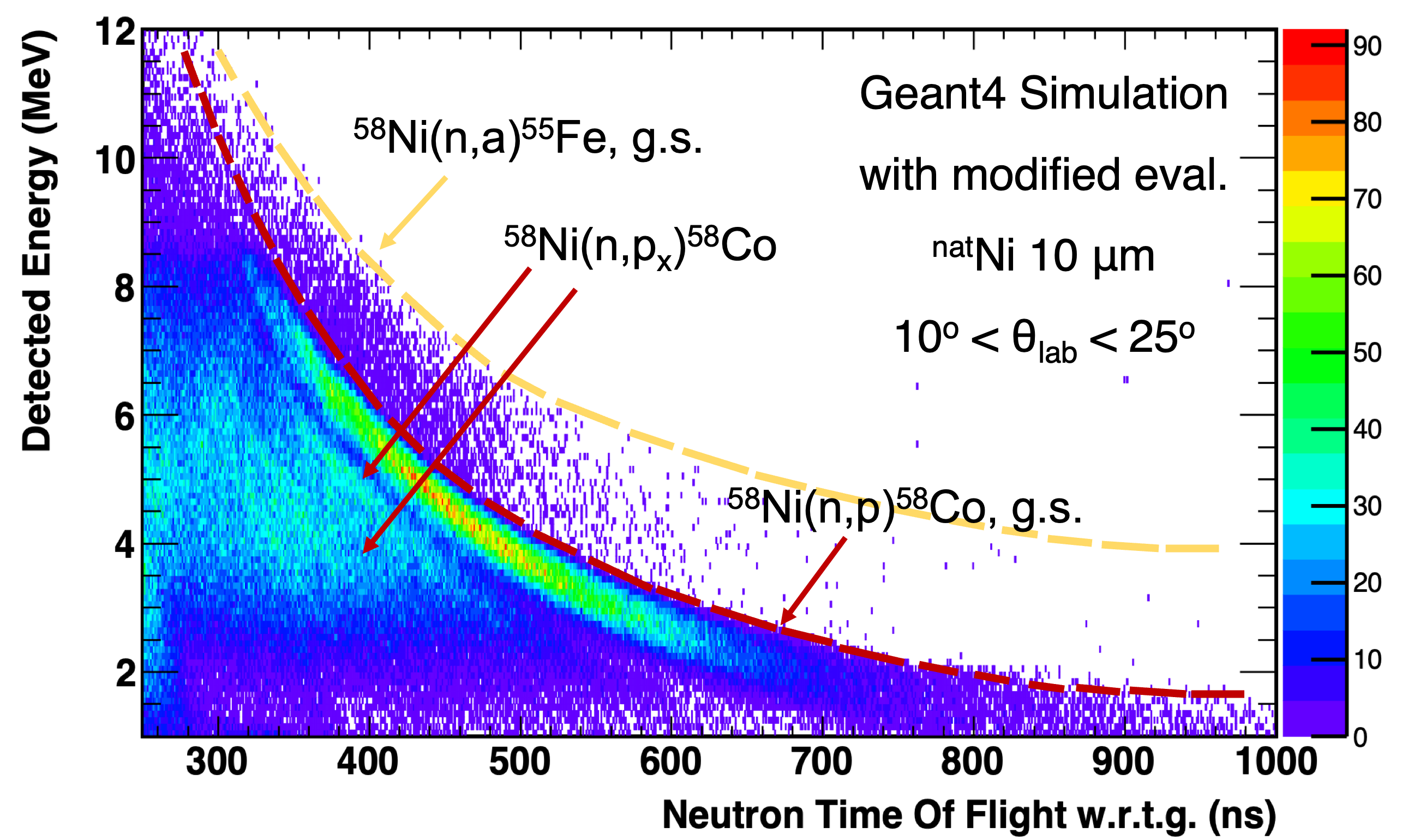}
\includegraphics[width=\linewidth]{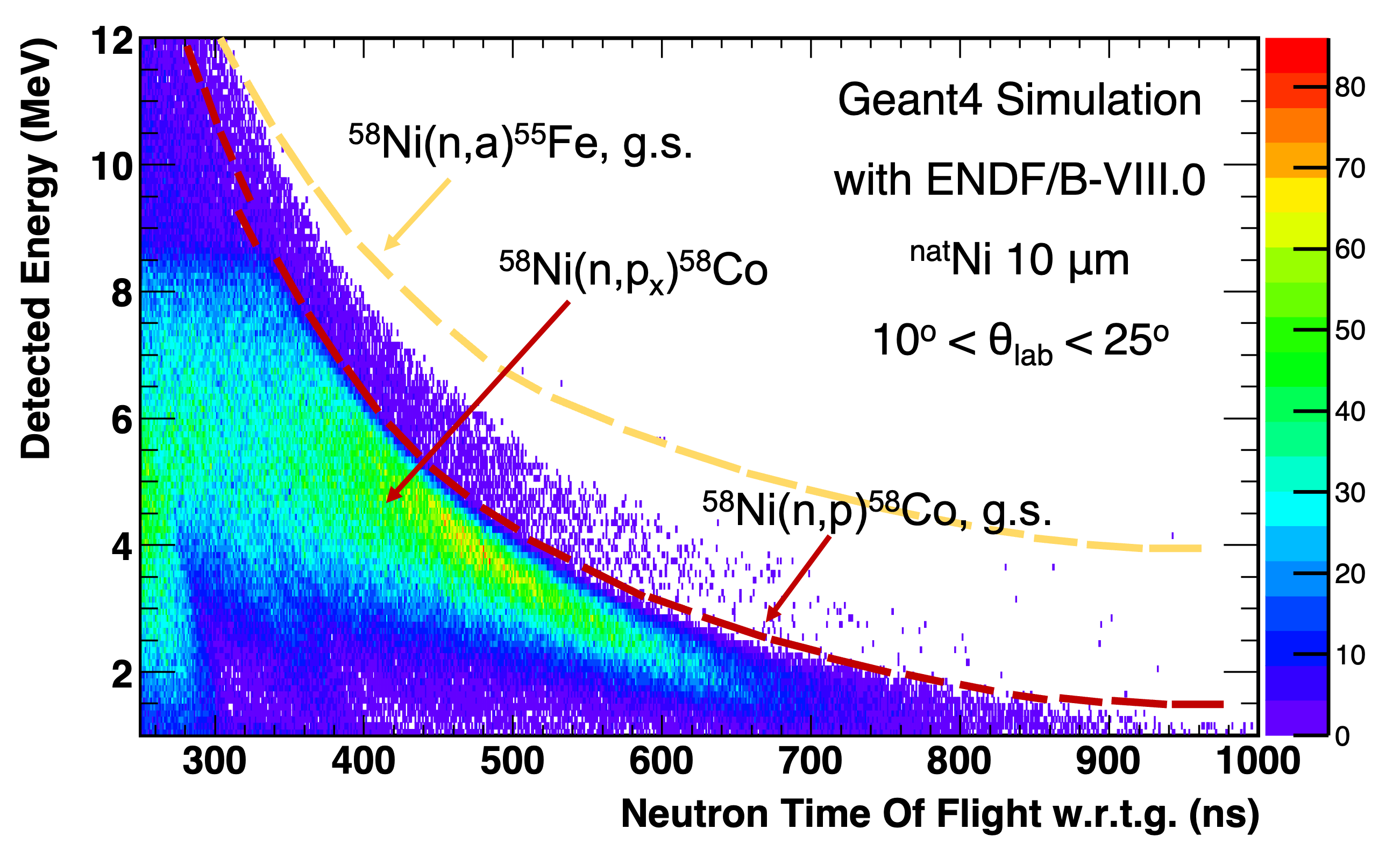}
\caption{Charged particle energy deposition as a function of incident neutron time of flight, in the detector at 4.1~cm using a 10~$\mu$m thick $^{nat}$Ni target (Top) Experiment, (Middle) GEANT4 simulation implementing a biasing factor of 500 and ENDF/B-VIII.0 data, and (Bottom) implementing the same biasing (500) but with data from new evaluation as an input.}
\label{natNi}
\end{figure}

The study of complex targets, such as the $^{nat}$Ni one that includes contributions from multiple isotopes, deems the existence of complete library input data sets crucial. $^{nat}$Ni consists of five different stable isotopes, with $^{58}$Ni being the most abundant one, comprising the 68$\%$ of natural nickel. The spectra from both the experiment and the GEANT4 simulation (implementing ENDF/B-VIII.0 data) are presented in Figure~\ref{natNi} (Top) and (Middle) respectively. Contrary to the excellent agreement shown in previous cases, the $^{nat}$Ni spectra appear to deviate considerably. Specifically, the experimental spectrum shows multiple excited states of protons from various (n,p$_x$) reaction channels on nickel isotopes, yet in the simulated one, these discrete levels are not independently visible as the shape of the outgoing charged particle spectra represents more of an averaged-out shape that is shifted towards lower energies.

To overcome this critical obstacle, a new, enriched evaluation for all nickel isotopes has been developed by a collaboration between KAERI and LANL.~\cite{KIM2020163699}. In this new evaluation, we included the partial cross-section data along with the angular distributions for the first forty excited states of the residual nuclei of each reaction and for every isotope of natural nickel. The application of this new, enhanced data set to our simulation raised some serious challenges, particularly in the incorporation of the angular distribution of every excitation level. Although the new library was accordingly modified to be implemented in the simulation, the corresponding Q-values involved (as discussed in section~\ref{sec:Geant_physics})  were assessed incorrectly by GEANT4 and required a slight modification of the relevant segment of source code in order to use the correct reaction Q-value and pass it correctly to the appropriate function. The spectrum generated with the corrected code is shown in Figure~\ref{natNi} (Bottom).

Comparing the outcome from the adjusted simulation Figure~\ref{natNi} (Bottom) to the experimental spectrum of Figure~\ref{natNi} (Top) we can observe a considerably better qualitative agreement, though still not completely satisfactory. The applied code adjustment successfully corrected the sampling of the reaction's Q-value for the particle emission leading to the ground state of the product nucleus, therefore the $^{58}$Ni(n,p$_0$)$^{58}$Co and $^{58}$Ni(n,a$_0$)$^{55}$Fe reaction channels are now clearly visible in the simulated spectrum Figure~\ref{natNi} (Bottom). However, the source code modification success was only partial since the mishandling of the (n,p$_x$) reaction channels with x$>$0  remains, as the bands corresponding to emission to excited states and that are expected to lie below the (n,p$_0$) ground state band are not as clearly distinguishable as in the experimental spectrum. This implies the need for further improvement of the GEANT4 code in order to appropriately handle based on differential cross-section data angular distributions of emitted particles that leave the product nucleus in an excited state.

\section{Conclusions}
\label{sec:Conclusions}

While the demand for high-quality nuclear data in modern applications is increasing, a new detector assembly, the Low Energy Neutron-induced charged-particle (Z) chamber (LENZ), has been developed at LANSCE. The LENZ chamber is conveniently designed for the study of neutron-induced charged particle reactions, over a broad neutron energy spectrum with a high total geometrical efficiency. In combination with the white neutron beam provided at LANSCE, the LENZ instrument covers an energy range from thermal energies up to several tens of MeV, which makes it ideal for double differential cross-section measurement studies. The absence of evaluated differential cross-section data for a large variety of medium and high-Z isotopes over a broad neutron energy range demonstrates one of the great potentials of the LENZ chamber and the measurements that can be performed using the detector assembly.
 
In the present work we developed a detailed simulation code for LENZ using the GEANT4 toolkit. The simulation consists of three main parts; namely the implementation of the white neutron beam as provided at the WNR facility, the tracking of neutron interactions with the target and the structural materials of the LENZ chamber, and the simulation of charged particle creation on the target and chamber materials along with their transport to the LENZ detector assembly. To ensure the reproduction of the reaction yields, the evaluated cross section and angular distribution data were obtained from the ENDF/B-VIII.0 evaluated nuclear data library. 

To overcome the inevitably low statistics associated with the generation of reaction products as secondary particles in the simulation in a reasonable amount of computing time, we implemented one of the standard GEANT4 biasing techniques to alter the frequency of occurrence of nuclear reactions in the materials studied. For consistency of the simulation, a validation test was performed in order to test the reliability of the biased simulation spectra. This test revealed that the maximum reliable biasing factor is inversely proportional to the thickness of the material on which the biasing is applied.

To verify the reliability of the simulated response of the detector system we performed a qualitative comparison test between selected experimental data and simulated results generated with the GEANT4 code. The results demonstrated a quite impressive qualitative agreement between the simulated and experimental spectra when the ENDF/B-VIII.0 evaluated library contained a complete set of angular distributions and cross-section data of the tested target material; in this case a Mylar and a Kapton foil. However, the limitations in the accuracy of the simulated results are revealed when the library is lacking partially or completely, accurate ($n,p_x$) and/or ($n,\alpha_x$) evaluated differential cross-section data. For the reference case of $^{nat}$Ni examined in this work, a new evaluation for all the nickel isotopes had to be developed at LANSCE in order to overcome this limitation. Parts of this evaluation are gradually being implemented in newer versions of the GEANT package by the developers. 

The implementation of the expanded data set revealed an issue in the handling of reaction Q-values in specific GEANT4 simulation code algorithms that affects the angular distribution of secondary particles. This part of the GEANT4 source code has received significant attention by the developers in more recent versions of the code and it is the authors' hope that future versions of GEANT4 will address it satisfactorily.

In conclusion, we have developed a detailed GEANT4 simulation code that reproduces the unique characteristics of the LENZ setup. We have benchmarked the simulation against a calculation with MCNP and validated  it using experimental data. The simulation code can now be used in combination with the LENZ setup during measurement planning or data analysis to  efficiently serve users in both fundamental nuclear physics research experiments, as well as applications of nuclear physics.

\section*{Acknowledgments}

This work benefited from the use of the LANSCE accelerator facility as was performed under the auspices of the National Nuclear Security Administration of the U.S. Department of Energy at Los Alamos National Laboratory under Contract No. 89233218CNA000001, the Laboratory Directed Research and Development program of Los Alamos National Laboratory under project number 20180228ER and the U.S.- ROK (Republic of Korea) International Nuclear Energy Research Initiative (INERI) Program of the U.S. Department of Energy's Office of Nuclear Energy under 2019-001-K. P. T. acknowledges support from Triad National Security, LLC under subcontract number 566247. This material is based upon work supported by the U.S. Department of Energy, Office of Science, Office of Nuclear Physics under Award Number DE-SC0014285.

\bibliography{NIMA,neutrino_p}

\begin{thebibliography}{47}
\expandafter\ifx\csname natexlab\endcsname\relax\def\natexlab#1{#1}\fi
\providecommand{\url}[1]{\texttt{#1}}
\providecommand{\href}[2]{#2}
\providecommand{\path}[1]{#1}
\providecommand{\DOIprefix}{doi:}
\providecommand{\ArXivprefix}{arXiv:}
\providecommand{\URLprefix}{URL: }
\providecommand{\Pubmedprefix}{pmid:}
\providecommand{\doi}[1]{\href{http://dx.doi.org/#1}{\path{#1}}}
\providecommand{\Pubmed}[1]{\href{pmid:#1}{\path{#1}}}
\providecommand{\bibinfo}[2]{#2}
\ifx\xfnm\relax \def\xfnm[#1]{\unskip,\space#1}\fi
\bibitem[{Burbidge et~al.(1957)Burbidge, Burbidge, Fowler, and Hoyle}]{b2fh}
\bibinfo{author}{E.~M. Burbidge}, \bibinfo{author}{G.~R. Burbidge},
  \bibinfo{author}{W.~A. Fowler}, \bibinfo{author}{F.~Hoyle},
\newblock \bibinfo{title}{Synthesis of the elements in stars},
\newblock \bibinfo{journal}{Rev. Mod. Phys.} \bibinfo{volume}{29}
  (\bibinfo{year}{1957}) \bibinfo{pages}{547--650}.
\bibitem[{Cameron(1957)}]{Cam57}
\bibinfo{author}{A.~Cameron},
\newblock \bibinfo{title}{Stellar evolution, nuclear astrophysics, and
  nucleogenesis},
\newblock \bibinfo{journal}{Chalk River Laboratory paper} \bibinfo{volume}{41}
  (\bibinfo{year}{1957}).
\bibitem[{Wallerstein et~al.(1997)Wallerstein, Iben, Parker, Boesgaard, Hale,
  Champagne, Barnes, K\"appeler, Smith, Hoffman, Timmes, Sneden, Boyd, Meyer,
  and Lambert}]{Wal97a}
\bibinfo{author}{G.~Wallerstein}, \bibinfo{author}{I.~Iben},
  \bibinfo{author}{P.~Parker}, \bibinfo{author}{A.~Boesgaard},
  \bibinfo{author}{G.~M. Hale}, \bibinfo{author}{A.~E. Champagne},
  \bibinfo{author}{C.~A. Barnes}, \bibinfo{author}{F.~K\"appeler},
  \bibinfo{author}{V.~Smith}, \bibinfo{author}{R.~Hoffman},
  \bibinfo{author}{F.~X. Timmes}, \bibinfo{author}{C.~Sneden},
  \bibinfo{author}{R.~N. Boyd}, \bibinfo{author}{B.~S. Meyer},
  \bibinfo{author}{D.~Lambert},
\newblock \bibinfo{title}{Synthesis of the elements in stars: forty years of
  progress},
\newblock \bibinfo{journal}{Rev. Mod. Phys.} \bibinfo{volume}{69}
  (\bibinfo{year}{1997}) \bibinfo{pages}{995--1084}.
\bibitem[{Frebel et~al.(2005)Frebel, Aoki, Christlieb, Ando, Asplund, Barklem,
  Beers, Eriksson, Fechner, Fujimoto, Honda, Kajino, Minezaki, Nomoto, Norris,
  Ryan, Takada-Hidai, Tsangarides, and Yoshii}]{Fre05a}
\bibinfo{author}{A.~Frebel}, \bibinfo{author}{W.~Aoki},
  \bibinfo{author}{N.~Christlieb}, \bibinfo{author}{H.~Ando},
  \bibinfo{author}{M.~Asplund}, \bibinfo{author}{P.~S. Barklem},
  \bibinfo{author}{T.~C. Beers}, \bibinfo{author}{K.~Eriksson},
  \bibinfo{author}{C.~Fechner}, \bibinfo{author}{M.~Y. Fujimoto},
  \bibinfo{author}{S.~Honda}, \bibinfo{author}{T.~Kajino},
  \bibinfo{author}{T.~Minezaki}, \bibinfo{author}{K.~Nomoto},
  \bibinfo{author}{J.~E. Norris}, \bibinfo{author}{S.~G. Ryan},
  \bibinfo{author}{M.~Takada-Hidai}, \bibinfo{author}{S.~Tsangarides},
  \bibinfo{author}{Y.~Yoshii},
\newblock \bibinfo{title}{Nucleosynthetic signatures of the first stars},
\newblock \bibinfo{journal}{Nature} \bibinfo{volume}{434}
  (\bibinfo{year}{2005}) \bibinfo{pages}{871--873}.
\bibitem[{Qian and Wasserburg(2007)}]{Qia07a}
\bibinfo{author}{Y.-Z. Qian}, \bibinfo{author}{G.~Wasserburg},
\newblock \bibinfo{title}{Where, oh where has the r-process gone?},
\newblock \bibinfo{journal}{Physics Reports} \bibinfo{volume}{442}
  (\bibinfo{year}{2007}) \bibinfo{pages}{237 -- 268}. \bibinfo{note}{The Hans
  Bethe Centennial Volume 1906-2006}.
\bibitem[{Montes et~al.(2007)Montes, Beers, Cowan, Elliot, Farouqi, Gallino,
  Heil, Kratz, Pfeiffer, Pignatari, and Schatz}]{Mon07a}
\bibinfo{author}{F.~Montes}, \bibinfo{author}{T.~C. Beers},
  \bibinfo{author}{J.~Cowan}, \bibinfo{author}{T.~Elliot},
  \bibinfo{author}{K.~Farouqi}, \bibinfo{author}{R.~Gallino},
  \bibinfo{author}{M.~Heil}, \bibinfo{author}{K.-L. Kratz},
  \bibinfo{author}{B.~Pfeiffer}, \bibinfo{author}{M.~Pignatari},
  \bibinfo{author}{H.~Schatz},
\newblock \bibinfo{title}{Nucleosynthesis in the early galaxy},
\newblock \bibinfo{journal}{The Astrophysical Journal} \bibinfo{volume}{671}
  (\bibinfo{year}{2007}) \bibinfo{pages}{1685}.
\bibitem[{Fr\"ohlich et~al.(2006)Fr\"ohlich, Mart\'inez-Pinedo, Liebend\"orfer,
  Thielemann, Bravo, Hix, Langanke, and Zinner}]{Fro06a}
\bibinfo{author}{C.~Fr\"ohlich}, \bibinfo{author}{G.~Mart\'inez-Pinedo},
  \bibinfo{author}{M.~Liebend\"orfer}, \bibinfo{author}{F.-K. Thielemann},
  \bibinfo{author}{E.~Bravo}, \bibinfo{author}{W.~R. Hix},
  \bibinfo{author}{K.~Langanke}, \bibinfo{author}{N.~T. Zinner},
\newblock \bibinfo{title}{Neutrino-induced nucleosynthesis of {A}$>$64 nuclei:
  The $\nu$p {P}rocess},
\newblock \bibinfo{journal}{Phys. Rev. Lett.} \bibinfo{volume}{96}
  (\bibinfo{year}{2006}) \bibinfo{pages}{142502}.
\bibitem[{Arcones and Montes(2011)}]{Arc11a}
\bibinfo{author}{A.~Arcones}, \bibinfo{author}{F.~Montes},
\newblock \bibinfo{title}{Production of light-element primary process nuclei in
  neutrino-driven winds},
\newblock \bibinfo{journal}{The Astrophysical Journal} \bibinfo{volume}{731}
  (\bibinfo{year}{2011}) \bibinfo{pages}{5}.
\bibitem[{{Liebend{\"o}rfer} et~al.(2003){Liebend{\"o}rfer}, {Mezzacappa},
  {Messer}, {Martinez-Pinedo}, {Hix}, and {Thielemann}}]{liebend03}
\bibinfo{author}{M.~{Liebend{\"o}rfer}}, \bibinfo{author}{A.~{Mezzacappa}},
  \bibinfo{author}{O.~E.~B. {Messer}}, \bibinfo{author}{G.~{Martinez-Pinedo}},
  \bibinfo{author}{W.~R. {Hix}}, \bibinfo{author}{F.-K. {Thielemann}},
\newblock \bibinfo{title}{{The neutrino signal in stellar core collapse and
  postbounce evolution}},
\newblock \bibinfo{journal}{Nuclear Physics A} \bibinfo{volume}{719}
  (\bibinfo{year}{2003}) \bibinfo{pages}{144--+}.
\bibitem[{{Buras} et~al.(2006){Buras}, {Rampp}, {Janka}, and
  {Kifonidis}}]{buras06b}
\bibinfo{author}{R.~{Buras}}, \bibinfo{author}{M.~{Rampp}},
  \bibinfo{author}{H.-T. {Janka}}, \bibinfo{author}{K.~{Kifonidis}},
\newblock \bibinfo{title}{{Two-dimensional hydrodynamic core-collapse supernova
  simulations with spectral neutrino transport. I. Numerical method and results
  for a 15 M{\.o} star}},
\newblock \bibinfo{journal}{{A}stronomy and {A}strophysics}
  \bibinfo{volume}{447} (\bibinfo{year}{2006}) \bibinfo{pages}{1049--1092}.
\bibitem[{{Fr{\"o}hlich} et~al.(2006){Fr{\"o}hlich}, {Hauser},
  {Liebend{\"o}rfer}, {Mart{\'{\i}}nez-Pinedo}, {Thielemann}, {Bravo},
  {Zinner}, {Hix}, {Langanke}, {Mezzacappa}, and {Nomoto}}]{cf06a}
\bibinfo{author}{C.~{Fr{\"o}hlich}}, \bibinfo{author}{P.~{Hauser}},
  \bibinfo{author}{M.~{Liebend{\"o}rfer}},
  \bibinfo{author}{G.~{Mart{\'{\i}}nez-Pinedo}}, \bibinfo{author}{F.-K.
  {Thielemann}}, \bibinfo{author}{E.~{Bravo}}, \bibinfo{author}{N.~T.
  {Zinner}}, \bibinfo{author}{W.~R. {Hix}}, \bibinfo{author}{K.~{Langanke}},
  \bibinfo{author}{A.~{Mezzacappa}}, \bibinfo{author}{K.~{Nomoto}},
\newblock \bibinfo{title}{{Composition of the Innermost Core-Collapse Supernova
  Ejecta}},
\newblock \bibinfo{journal}{The Astrophysical Journal} \bibinfo{volume}{637}
  (\bibinfo{year}{2006}) \bibinfo{pages}{415--426}.
\bibitem[{{Fischer} et~al.(2010){Fischer}, {Whitehouse}, {Mezzacappa},
  {Thielemann}, and {Liebend{\"o}rfer}}]{fischer10}
\bibinfo{author}{T.~{Fischer}}, \bibinfo{author}{S.~C. {Whitehouse}},
  \bibinfo{author}{A.~{Mezzacappa}}, \bibinfo{author}{F.-K. {Thielemann}},
  \bibinfo{author}{M.~{Liebend{\"o}rfer}},
\newblock \bibinfo{title}{{Protoneutron star evolution and the neutrino-driven
  wind in general relativistic neutrino radiation hydrodynamics simulations}},
\newblock \bibinfo{journal}{{A}stronomy and {A}strophysics}
  \bibinfo{volume}{517} (\bibinfo{year}{2010}) \bibinfo{pages}{A80+}.
\bibitem[{{H{\"u}depohl} et~al.(2010){H{\"u}depohl}, {M{\"u}ller}, {Janka},
  {Marek}, and {Raffelt}}]{huedepohl10}
\bibinfo{author}{L.~{H{\"u}depohl}}, \bibinfo{author}{B.~{M{\"u}ller}},
  \bibinfo{author}{H.-T. {Janka}}, \bibinfo{author}{A.~{Marek}},
  \bibinfo{author}{G.~G. {Raffelt}},
\newblock \bibinfo{title}{{Neutrino Signal of Electron-Capture Supernovae from
  Core Collapse to Cooling}},
\newblock \bibinfo{journal}{Physical Review Letters} \bibinfo{volume}{104}
  (\bibinfo{year}{2010}) \bibinfo{pages}{251101}.
\bibitem[{{Pruet} et~al.(2006){Pruet}, {Hoffman}, {Woosley}, {Janka}, and
  {Buras}}]{pruetII}
\bibinfo{author}{J.~{Pruet}}, \bibinfo{author}{R.~D. {Hoffman}},
  \bibinfo{author}{S.~E. {Woosley}}, \bibinfo{author}{H.-T. {Janka}},
  \bibinfo{author}{R.~{Buras}},
\newblock \bibinfo{title}{{Nucleosynthesis in Early Supernova Winds. II. The
  Role of Neutrinos}},
\newblock \bibinfo{journal}{The Astrophysical Journal} \bibinfo{volume}{644}
  (\bibinfo{year}{2006}) \bibinfo{pages}{1028--1039}.
\bibitem[{{Wanajo}(2006)}]{wanajo06}
\bibinfo{author}{S.~{Wanajo}},
\newblock \bibinfo{title}{{The rp-Process in Neutrino-driven Winds}},
\newblock \bibinfo{journal}{The Astrophysical Journal} \bibinfo{volume}{647}
  (\bibinfo{year}{2006}) \bibinfo{pages}{1323--1340}.
\bibitem[{Wanajo et~al.(2011)Wanajo, Janka, and Kubono}]{Wan11a}
\bibinfo{author}{S.~Wanajo}, \bibinfo{author}{H.-T. Janka},
  \bibinfo{author}{S.~Kubono},
\newblock \bibinfo{title}{Uncertainties in the $\nu$p-process: Supernova
  dynamics versus nuclear physics},
\newblock \bibinfo{journal}{The Astrophysical Journal} \bibinfo{volume}{729}
  (\bibinfo{year}{2011}) \bibinfo{pages}{46}.
\bibitem[{{Arcones} et~al.(2012){Arcones}, {Fr\"ohlich}, and
  {Mart{\'{\i}}nez-Pinedo}}]{aacf}
\bibinfo{author}{A.~{Arcones}}, \bibinfo{author}{C.~{Fr\"ohlich}},
  \bibinfo{author}{G.~{Mart{\'{\i}}nez-Pinedo}},
\newblock \bibinfo{title}{{Impact of Supernova Dynamics on the
  {$\nu$}p-process}},
\newblock \bibinfo{journal}{The Astrophysical Journal} \bibinfo{volume}{750}
  (\bibinfo{year}{2012}) \bibinfo{pages}{18}.
\bibitem[{Gastis et~al.(2020)Gastis, Perdikakis, Dissanayake, Tsintari,
  Sultana, Brune, Massey, Meisel, Voinov, Brandenburg, Danley, Giri,
  Jones-Alberty, Paneru, Soltesz, and Subedi}]{Gastis2020a}
\bibinfo{author}{P.~Gastis}, \bibinfo{author}{G.~Perdikakis},
  \bibinfo{author}{J.~Dissanayake}, \bibinfo{author}{P.~Tsintari},
  \bibinfo{author}{I.~Sultana}, \bibinfo{author}{C.~R. Brune},
  \bibinfo{author}{T.~N. Massey}, \bibinfo{author}{Z.~Meisel},
  \bibinfo{author}{A.~V. Voinov}, \bibinfo{author}{K.~Brandenburg},
  \bibinfo{author}{T.~Danley}, \bibinfo{author}{R.~Giri},
  \bibinfo{author}{Y.~Jones-Alberty}, \bibinfo{author}{S.~Paneru},
  \bibinfo{author}{D.~Soltesz}, \bibinfo{author}{S.~Subedi},
\newblock \bibinfo{title}{Constraining the destruction rate of
  $^{40}\mathrm{K}$ in stellar nucleosynthesis through the study of the
  $^{40}\mathrm{Ar}(p,n)^{40}\mathrm{K}$ reaction},
\newblock \bibinfo{journal}{Phys. Rev. C} \bibinfo{volume}{101}
  (\bibinfo{year}{2020}) \bibinfo{pages}{055805}.
\bibitem[{Mazzone et~al.(2018)Mazzone, Finocchiaro, Lo Meo, and
  Colonna}]{MAZZONE201833}
\bibinfo{author}{A.~Mazzone}, \bibinfo{author}{P.~Finocchiaro},
  \bibinfo{author}{S.~Lo Meo}, \bibinfo{author}{N.~Colonna},
\newblock \bibinfo{title}{{GEANT4} simulations of a novel  $^3${H}e-free
  thermalization neutron detector},
\newblock \bibinfo{journal}{Nuclear Instruments and Methods in Physics Research
  Section A: Accelerators, Spectrometers, Detectors and Associated Equipment}
  \bibinfo{volume}{889} (\bibinfo{year}{2018}) \bibinfo{pages}{33 -- 38}.
\bibitem[{Lalremruata et~al.(2012)Lalremruata, Otuka, Tambave, Mulik, Patil,
  Dhole, Saxena, Ganesan, and Bhoraskar}]{PhysRevC.85.024624}
\bibinfo{author}{B.~Lalremruata}, \bibinfo{author}{N.~Otuka},
  \bibinfo{author}{G.~J. Tambave}, \bibinfo{author}{V.~K. Mulik},
  \bibinfo{author}{B.~J. Patil}, \bibinfo{author}{S.~D. Dhole},
  \bibinfo{author}{A.~Saxena}, \bibinfo{author}{S.~Ganesan},
  \bibinfo{author}{V.~N. Bhoraskar},
\newblock \bibinfo{title}{Systematic study of (n,p) reaction cross sections
  from the reaction threshold to 20 {M}e{V}},
\newblock \bibinfo{journal}{Phys. Rev. C} \bibinfo{volume}{85}
  (\bibinfo{year}{2012}) \bibinfo{pages}{024624}.
\bibitem[{Gastis et~al.(2021)Gastis, Perdikakis, Berg, Dombos, Estrade,
  Falduto, Horoi, Liddick, Lipschutz, Lyons, Montes, Palmisano, Pereira,
  Randhawa, Redpath, Redshaw, Schmitt, Sheehan, Smith, Tsintari, Villari, Wang,
  and Zegers}]{Gastis2021}
\bibinfo{author}{P.~Gastis}, \bibinfo{author}{G.~Perdikakis},
  \bibinfo{author}{G.~P.~A. Berg}, \bibinfo{author}{A.~C. Dombos},
  \bibinfo{author}{A.~Estrade}, \bibinfo{author}{A.~Falduto},
  \bibinfo{author}{M.~Horoi}, \bibinfo{author}{S.~N. Liddick},
  \bibinfo{author}{S.~Lipschutz}, \bibinfo{author}{S.~Lyons},
  \bibinfo{author}{F.~Montes}, \bibinfo{author}{A.~Palmisano},
  \bibinfo{author}{J.~Pereira}, \bibinfo{author}{J.~S. Randhawa},
  \bibinfo{author}{T.~Redpath}, \bibinfo{author}{M.~Redshaw},
  \bibinfo{author}{J.~Schmitt}, \bibinfo{author}{J.~Sheehan},
  \bibinfo{author}{M.~K. Smith}, \bibinfo{author}{P.~Tsintari},
  \bibinfo{author}{A.~C.~C. Villari}, \bibinfo{author}{K.~Wang},
  \bibinfo{author}{R.~G.~T. Zegers},
\newblock \bibinfo{title}{A technique for the study of (p,n) reactions with
  unstable isotopes at energies relevant to astrophysics},
\newblock \bibinfo{journal}{Nuclear Instruments \& Methods in Physics Research
  Section A-accelerators Spectrometers Detectors and Associated Equipment}
  \bibinfo{volume}{985} (\bibinfo{year}{2021}) \bibinfo{pages}{164603}.
\bibitem[{Lee et~al.(tted)Lee, Mosby, Prokop, Long, G{\"{o}}rres, Stech, and
  Wiescher}]{LeePSD2019}
\bibinfo{author}{H.~Y. Lee}, \bibinfo{author}{S.~Mosby},
  \bibinfo{author}{C.~Prokop}, \bibinfo{author}{A.~Long},
  \bibinfo{author}{J.~G{\"{o}}rres}, \bibinfo{author}{E.~Stech},
  \bibinfo{author}{M.~Wiescher},
\newblock \bibinfo{title}{Low {E}nergy {N}eutron-induced {C}harged-particle
  ({Z}) ({LENZ}) instrument development with a focus on the {P}ulse {S}hape
  {D}iscrimination for double-sided silicon strip detectors at {LANSCE}},
\newblock \bibinfo{journal}{Nuclear Instruments and Methods in Physics Research
  Section A: Accelerators, Spectrometers, Detectors and Associated Equipment}
  (\bibinfo{year}{submitted}).
\bibitem[{Haight(2008)}]{NZhaight}
\bibinfo{author}{R.~C. Haight},
\newblock \bibinfo{title}{Hydrogen and helium production in structural
  materials by neutrons},
\newblock \bibinfo{journal}{EPJ Web of Conferences}  (\bibinfo{year}{2008})
  \bibinfo{pages}{1081 -- 1084}. \bibinfo{note}{Proc. Int. Conf. on Nuclear
  Data for Science and Technology, 22 -- 27 Apr., 2007, Nice, France, Ed. O.
  Bersillon, F. Gunsing, E. Bauge, R. Jacqmin, and S. Leray}.
\bibitem[{Devlin et~al.(2009)Devlin, Taddeucci, Hale, Haight, and
  O'Donnell}]{Devlin2009}
\bibinfo{author}{M.~Devlin}, \bibinfo{author}{T.~N. Taddeucci},
  \bibinfo{author}{G.~M. Hale}, \bibinfo{author}{R.~C. Haight},
  \bibinfo{author}{J.~O'Donnell},
\newblock \bibinfo{title}{Differential cross section measurements for the
  ${}^6$\uppercase{L}i(n,t)$\alpha$ reaction in the few
  \uppercase{M}e\uppercase{V} region},
\newblock \bibinfo{journal}{Proceeding In Capture Gamma-Ray Spectroscopy and
  Related Topics} \bibinfo{volume}{1090} (\bibinfo{year}{2009})
  \bibinfo{pages}{215}.
\bibitem[{Kunieda et~al.(2012)Kunieda, Haight, Kawano, Chadwick, Sterbenz,
  Bateman, Wasson, Grimes, Maier-Komor, Vonach, Fukahori, and
  Watanabe}]{Kunieda2012}
\bibinfo{author}{S.~Kunieda}, \bibinfo{author}{R.~C. Haight},
  \bibinfo{author}{T.~Kawano}, \bibinfo{author}{M.~B. Chadwick},
  \bibinfo{author}{S.~M. Sterbenz}, \bibinfo{author}{F.~B. Bateman},
  \bibinfo{author}{O.~A. Wasson}, \bibinfo{author}{S.~M. Grimes},
  \bibinfo{author}{P.~Maier-Komor}, \bibinfo{author}{H.~Vonach},
  \bibinfo{author}{T.~Fukahori}, \bibinfo{author}{Y.~Watanabe},
\newblock \bibinfo{title}{Measurement and model analysis of
  $(n,x\ensuremath{\alpha})$ cross sections for \uppercase{C}r, \uppercase{F}e,
  ${}^{59}$\uppercase{C}o, and ${}^{58,60}$\uppercase{N}i from threshold energy
  to 150~\uppercase{M}e\uppercase{V}},
\newblock \bibinfo{journal}{Phys. Rev. C} \bibinfo{volume}{85}
  (\bibinfo{year}{2012}) \bibinfo{pages}{054602}.
\bibitem[{Haight et~al.(1994)Haight, Lee, and Sterbenz}]{osti_10165734}
\bibinfo{author}{R.~Haight}, \bibinfo{author}{T.~Lee},
  \bibinfo{author}{S.~Sterbenz},
\newblock \bibinfo{title}{Neutron-induced charged-particle emission studies
  below 100 {M}e{V} at {WNR}}  (\bibinfo{year}{1994}).
\bibitem[{Allison and \textit{et al.}(2003)}]{AGOSTINELLI2003250}
\bibinfo{author}{J.~Allison}, \bibinfo{author}{\textit{et al.}},
\newblock \bibinfo{title}{Geant4 - {A} simulation toolkit},
\newblock \bibinfo{journal}{Nuclear Instruments and Methods in Physics Research
  Section A: Accelerators, Spectrometers, Detectors and Associated Equipment}
  \bibinfo{volume}{506} (\bibinfo{year}{2003}) \bibinfo{pages}{250 -- 303}.
\bibitem[{\v{Z}ugec and \textit{et al.}(2014)}]{ZUGEC201457}
\bibinfo{author}{P.~\v{Z}ugec}, \bibinfo{author}{\textit{et al.}},
\newblock \bibinfo{title}{{GEANT4} simulation of the neutron background of the
  {C6D6} set-up for capture studies at n{TOF}},
\newblock \bibinfo{journal}{Nuclear Instruments and Methods in Physics Research
  Section A: Accelerators, Spectrometers, Detectors and Associated Equipment}
  \bibinfo{volume}{760} (\bibinfo{year}{2014}) \bibinfo{pages}{57 -- 67}.
\bibitem[{Garcia et~al.(2017)Garcia, Mendoza, Cano-Ott, Nolte, Martinez,
  Algora, Tain, Banerjee, and Bhattacharya}]{GARCIA201773}
\bibinfo{author}{A.~Garcia}, \bibinfo{author}{E.~Mendoza},
  \bibinfo{author}{D.~Cano-Ott}, \bibinfo{author}{R.~Nolte},
  \bibinfo{author}{T.~Martinez}, \bibinfo{author}{A.~Algora},
  \bibinfo{author}{J.~Tain}, \bibinfo{author}{K.~Banerjee},
  \bibinfo{author}{C.~Bhattacharya},
\newblock \bibinfo{title}{New physics model in geant4 for the simulation of
  neutron interactions with organic scintillation detectors},
\newblock \bibinfo{journal}{Nuclear Instruments and Methods in Physics Research
  Section A: Accelerators, Spectrometers, Detectors and Associated Equipment}
  \bibinfo{volume}{868} (\bibinfo{year}{2017}) \bibinfo{pages}{73 -- 81}.
\bibitem[{Kim et~al.(2020)Kim, Lee, Kawano, Georgiadou, Kuvin, Zavorka, and
  Herman}]{KIM2020163699}
\bibinfo{author}{H.~I. Kim}, \bibinfo{author}{H.~Y. Lee},
  \bibinfo{author}{T.~Kawano}, \bibinfo{author}{A.~Georgiadou},
  \bibinfo{author}{S.~A. Kuvin}, \bibinfo{author}{L.~Zavorka},
  \bibinfo{author}{M.~W. Herman},
\newblock \bibinfo{title}{New evaluation on angular distributions and energy
  spectra for neutron-induced charged-particle measurements},
\newblock \bibinfo{journal}{Nuclear Instruments and Methods in Physics Research
  Section A: Accelerators, Spectrometers, Detectors and Associated Equipment}
  \bibinfo{volume}{963} (\bibinfo{year}{2020}) \bibinfo{pages}{163699}.
\bibitem[{Lisowski et~al.(1990)Lisowski, Bowman, Russell, and
  Wender}]{Lisowski1990}
\bibinfo{author}{P.~W. Lisowski}, \bibinfo{author}{C.~D. Bowman},
  \bibinfo{author}{G.~J. Russell}, \bibinfo{author}{S.~A. Wender},
\newblock \bibinfo{title}{The {L}os {A}lamos {N}ational {L}aboratory spallation
  neutron sources},
\newblock \bibinfo{journal}{Nuclear Science and Engineering}
  \bibinfo{volume}{106} (\bibinfo{year}{1990}) \bibinfo{pages}{208--218}.
\bibitem[{Lisowski and Schoenberg(2006)}]{Lisowski2006}
\bibinfo{author}{P.~W. Lisowski}, \bibinfo{author}{K.~F. Schoenberg},
\newblock \bibinfo{title}{The {L}os {A}lamos {N}eutron {S}cience {C}enter},
\newblock \bibinfo{journal}{Nuclear Instruments and Methods in Physics Research
  Section A: Accelerators, Spectrometers, Detectors and Associated Equipment}
  \bibinfo{volume}{562} (\bibinfo{year}{2006}) \bibinfo{pages}{910--914}.
\bibitem[{LAN(2020)}]{LANSCE_fp}
\bibinfo{title}{L{A}{N}{S}{C}{E} flight paths description},
  \bibinfo{year}{2020}. \URLprefix
  \url{https://lansce.lanl.gov/facilities/wnr/flight-paths/index.php}.
\bibitem[{Werner(2017)}]{Werner2017}
\bibinfo{author}{C.~Werner},
\newblock \bibinfo{title}{{MCNP}\textsuperscript{\textcopyright} {U}ser's
  {M}anual, {C}ode {V}ersion 6.2, {LA-UR}-17-29981}  (\bibinfo{year}{2017}).
\bibitem[{Lee et~al.(tion)Lee, Kuvin, Zavorka, Hale, and Paris}]{Lee_O16_2020}
\bibinfo{author}{H.~Y. Lee}, \bibinfo{author}{S.~A. Kuvin},
  \bibinfo{author}{L.~Zavorka}, \bibinfo{author}{G.~Hale},
  \bibinfo{author}{M.~Paris},
\newblock \bibinfo{title}{New high-precision measurement on
  ${}^{16}$\uppercase{O}(n,$\alpha$)${}^{13}$\uppercase{C} reaction using a
  white neutron source},
\newblock \bibinfo{journal}{Phys. Rev. C}  (\bibinfo{year}{in preparation}).
\bibitem[{Kuvin et~al.(ress)Kuvin, Lee, Kawano, Di{G}iovine, Georgiadou,
  Vermeulen, White, Zavorka, and Kim}]{Kuvin_Cl35_2020}
\bibinfo{author}{S.~A. Kuvin}, \bibinfo{author}{H.~Y. Lee},
  \bibinfo{author}{T.~Kawano}, \bibinfo{author}{B.~Di{G}iovine},
  \bibinfo{author}{A.~Georgiadou}, \bibinfo{author}{C.~Vermeulen},
  \bibinfo{author}{M.~White}, \bibinfo{author}{L.~Zavorka},
  \bibinfo{author}{H.~I. Kim},
\newblock \bibinfo{title}{Non-statistical fluctuations in the
  ${}^{35}$\uppercase{C}l(n,p)${}^{35}$\uppercase{S} reaction for fast neutron
  energies},
\newblock \bibinfo{journal}{Phys. Rev. C}  (\bibinfo{year}{in press}).
\bibitem[{Mic(2020)}]{Micron}
\bibinfo{title}{Micron {S}emiconductor {L}td.}, \bibinfo{year}{2020}.
  \URLprefix \url{http://www.micronsemiconductor.co.uk/}.
\bibitem[{Civ(2020)}]{Cividec}
\bibinfo{title}{C{I}{V}{I}{D}{E}{C} {I}nstrumentation}, \bibinfo{year}{2020}.
  \URLprefix \url{http://www.cividec.at}.
\bibitem[{Wright and \textit{et al.}(2016)}]{ALLISON2016186}
\bibinfo{author}{D.~Wright}, \bibinfo{author}{\textit{et al.}},
\newblock \bibinfo{title}{Recent developments in {G}eant4},
\newblock \bibinfo{journal}{Nuclear Instruments and Methods in Physics Research
  Section A: Accelerators, Spectrometers, Detectors and Associated Equipment}
  \bibinfo{volume}{835} (\bibinfo{year}{2016}) \bibinfo{pages}{186 -- 225}.
\bibitem[{{Allison} and \textit{et al.}(2006)}]{1610988}
\bibinfo{author}{J.~{Allison}}, \bibinfo{author}{\textit{et al.}},
\newblock \bibinfo{title}{Geant4 developments and applications},
\newblock \bibinfo{journal}{IEEE Transactions on Nuclear Science}
  \bibinfo{volume}{53} (\bibinfo{year}{2006}) \bibinfo{pages}{270--278}.
\bibitem[{Ivanchenko et~al.(2017)Ivanchenko, Dondero, Fioretti, Ivantchenko,
  Lei, Lotti, Mantero, and Mineo}]{Ivanchenko2017}
\bibinfo{author}{V.~Ivanchenko}, \bibinfo{author}{P.~Dondero},
  \bibinfo{author}{V.~Fioretti}, \bibinfo{author}{A.~Ivantchenko},
  \bibinfo{author}{F.~Lei}, \bibinfo{author}{S.~Lotti},
  \bibinfo{author}{A.~Mantero}, \bibinfo{author}{T.~Mineo},
\newblock \bibinfo{title}{Validation of {G}eant4 10.3 simulation of proton
  interaction for space radiation effects},
\newblock \bibinfo{journal}{Experimental Astronomy} \bibinfo{volume}{44}
  (\bibinfo{year}{2017}) \bibinfo{pages}{437--450}.
\bibitem[{Vagena et~al.(2020)Vagena, Androulakaki, Kokkoris, Patronis, and
  Stamati}]{VAGENA202044}
\bibinfo{author}{E.~Vagena}, \bibinfo{author}{E.~Androulakaki},
  \bibinfo{author}{M.~Kokkoris}, \bibinfo{author}{N.~Patronis},
  \bibinfo{author}{M.~Stamati},
\newblock \bibinfo{title}{A comparative study of stopping power calculations
  implemented in monte carlo codes and compilations with experimental data},
\newblock \bibinfo{journal}{Nuclear Instruments and Methods in Physics Research
  Section B: Beam Interactions with Materials and Atoms} \bibinfo{volume}{467}
  (\bibinfo{year}{2020}) \bibinfo{pages}{44 -- 52}.
\bibitem[{Mendoza et~al.(2018)Mendoza, Cano-Ott, and Capote}]{ENDFVIII_2018}
\bibinfo{author}{E.~Mendoza}, \bibinfo{author}{D.~Cano-Ott},
  \bibinfo{author}{R.~Capote},
\newblock \bibinfo{title}{Update of the {E}valuated {N}eutron {C}ross {S}ection
  {L}ibraries for the {G}eant4 code},
\newblock \bibinfo{journal}{IAEA technical report INDC (NDS)}
  \bibinfo{volume}{-0758} (\bibinfo{year}{2018}).
\bibitem[{Mendoza et~al.(2012)Mendoza, Cano-Ott, Guerrero, and
  Capote}]{ENDFVIII_2012}
\bibinfo{author}{E.~Mendoza}, \bibinfo{author}{D.~Cano-Ott},
  \bibinfo{author}{D.~Guerrero}, \bibinfo{author}{R.~Capote},
\newblock \bibinfo{title}{New evaluated neutron cross section libraries for the
  {G}eant4 code},
\newblock \bibinfo{journal}{IAEA technical report INDC (NDS)}
  \bibinfo{volume}{-0612} (\bibinfo{year}{2012}).
\bibitem[{Mendoza et~al.(2014)Mendoza, Cano-Ott, Koi, and
  Guerrero}]{Mendoza20142357}
\bibinfo{author}{E.~Mendoza}, \bibinfo{author}{D.~Cano-Ott},
  \bibinfo{author}{T.~Koi}, \bibinfo{author}{C.~Guerrero},
\newblock \bibinfo{title}{New standard evaluated neutron cross section
  libraries for the geant4 code and first verification},
\newblock \bibinfo{journal}{IEEE Transactions on Nuclear Science}
  \bibinfo{volume}{61} (\bibinfo{year}{2014}) \bibinfo{pages}{2357--2364}.
\bibitem[{{Ribon} et~al.(2009){Ribon}, {Apostolakis}, {Dotti}, {Folger},
  {Grichine}, {Ivanchenko}, {Kosov}, {Uzhinsky}, and {Wright}}]{HadronModels}
\bibinfo{author}{A.~{Ribon}}, \bibinfo{author}{J.~{Apostolakis}},
  \bibinfo{author}{A.~{Dotti}}, \bibinfo{author}{G.~{Folger}},
  \bibinfo{author}{V.~{Grichine}}, \bibinfo{author}{V.~{Ivanchenko}},
  \bibinfo{author}{M.~{Kosov}}, \bibinfo{author}{V.~{Uzhinsky}},
  \bibinfo{author}{D.~H. {Wright}},
\newblock \bibinfo{title}{Transition between hadronic models in geant4},
\newblock in: \bibinfo{booktitle}{2009 IEEE Nuclear Science Symposium
  Conference Record (NSS/MIC)}, \bibinfo{year}{2009}, pp.
  \bibinfo{pages}{526--529}.
\bibitem[{Wright and Kelsey(2015)}]{Bertini}
\bibinfo{author}{D.~Wright}, \bibinfo{author}{M.~Kelsey},
\newblock \bibinfo{title}{The geant4 bertini cascade},
\newblock \bibinfo{journal}{Nuclear Instruments and Methods in Physics Research
  Section A: Accelerators, Spectrometers, Detectors and Associated Equipment}
  \bibinfo{volume}{804} (\bibinfo{year}{2015}) \bibinfo{pages}{175 -- 188}.

\end{thebibliography}

\end{document}